\documentclass[11pt]{preprint}
\usepackage{amsmath,amssymb,epsfig}
\usepackage{times}
\newtheorem{theorem}{Theorem}[section]
\newtheorem{proposition}[theorem]{Proposition}

\newtheorem{lemma}[theorem]{Lemma}

\newtheorem{remark}[theorem]{Remark}
\def\l{\ell}
\let\kappa=\varkappa
\makeatletter
\def\SI#1{\settowidth{\labelwidth}{#1}%
        \leftmargini\labelwidth
        \advance\leftmargini\labelsep
   \def\@listi{\leftmargin\leftmargini
        \labelwidth\leftmargini\advance\labelwidth by -\labelsep
        \parsep \lineskip
        \topsep  6\p@ \@plus2\p@ \@minus4\p@
        \itemsep\parsep}}

\renewcommand{\theequation}{\thesection.\arabic{equation}}
\renewcommand{\subsection}{\@startsection{subsection}{2}{0mm}{-\baselineskip}{0.5\baselineskip}{\normalfont\large\bf}}
\renewcommand{\subsubsection}{\@startsection{subsubsection}{2}{0mm}{-\baselineskip}{0.5\baselineskip}{\normalfont\normalsize\it}}
\renewcommand{\@oddfoot}{}%
\def\blfootnote{\xdef\@thefnmark{}\@footnotetext}
\makeatother

\def\dlap#1{{\setbox0=\hbox{#1}\dimen0=\ht0\divide\dimen0by
1\raise-\dimen0\box0}}
\def\ulap#1{{\setbox0=\hbox{#1}\dimen0
=\ht0\divide\dimen0
by 1\raise\dimen0\box0}}
\def\nlap#1{{\setbox0=\hbox{#1}\dimen0
=\ht0\divide\dimen0
by 2\raise-\dimen0\box0}}
\def\figureeewithtex#1#2#3#4#5#6#7#8#9{
\def\myunit{1cm}
\setlength{\unitlength}{\myunit}
\def\mywidth{#1}
\def\myheight{#2}
\def\myxorig{7.5}
\def\myyorig{5}
\def\picoffset{#3}
\def\mytxoff{#4}
\def\mytyoff{#5}

\begin{figure}\label{fig:#9}
\C{\psfig{figure=#6}}%
\leavevmode
\input{#7}
\end{picture}%
\null\vskip\picoffset cm
\caption{#8}
\end{figure}
}%

\def\figurewithtex#1#2#3#4#5#6#7#8#9{
\def\myunit{1cm}
\setlength{\unitlength}{\myunit}
\def\mywidth{#1}
\def\myheight{#2}
\def\myxorig{7.5}
\def\myyorig{5}
\def\picoffset{#3}
\def\mytxoff{#4}
\def\mytyoff{#5}

\begin{figure}[ht]
\leavevmode\raise0\unitlength\vbox to\myheight\unitlength{%
\vfill\hbox to\mywidth\unitlength{%
\hfill%
\newdimen\figcenter
\figcenter=\hsize\relax
\divide\figcenter by 2%
\begin{picture}(-15,0)(\myxorig,\myyorig)%
\put(0,0){\makebox(21,10)[lb]{\epsfig{figure=#6}}}%
\input{#7}
\end{picture}%
\hfill}%
\vfill}%
\null\vskip\picoffset cm
\caption{#8}\label{#9}
\end{figure}
}%

\usepackage{sub_JP}
\usepackage{MHequ}
\usepackage{ScriptFonts}
\usepackage{myarray}
\usepackage{Symbols}
\def\LTE/{local thermal equilibrium}
\def\L{{\rm L}}
\def\R{{\rm R}}
\def\gammal{\gamma_{\rm L}}
\def\gammar{\gamma_{\rm R}}
\def\tl{T_{\rm L}}
\def\tr{T_{\rm R}}
\def\rate{\rho}
\def\jl{\rate_{\rm L}}
\def\jr{\rate_{\rm R}}
\def\j{\rate}
\def\Tr{T,\rate}

\let\rho=\varrho
\let\epsilon=\varepsilon
\let\theta=\vartheta
\def\oo{{\it o}}
\def\OO{{\cal O}}
\def\ie{{\it i.e.}}

\def\x{6.2cm}
\def\C#1{\begin{center}{#1}\end{center}}
\let\psfig=\epsfig
\def\Pr{{\bf P}}
\mathchardef\emptyset="023B%

\begin{document}
\title{Nonequilibrium Energy Profiles for a Class of 1-D Models}
\author{Jean-Pierre Eckmann${}^{1,2}$ and Lai-Sang 
Young${}^3$}
\institute{${}^1$D\'epartement de Physique Th\'eorique, Universit\'e de
Gen\`eve,\\${}^2$Section de Math\'ematiques, Universit\'e de
Gen\`eve,\\
${}^3$Courant Institute for Mathematical Sciences, New York
University}
\maketitle

\blfootnote{JPE is partially supported by the 
Fonds National Suisse.}
\blfootnote{LSY is partially supported by NSF Grant \#0100538.}

\bigskip
\begin{abstract}
As a paradigm for heat conduction in 1 dimension, we propose
a class of models represented by chains of identical cells,
each one of which containing an energy storage device called a ``tank". 
Energy exchange among tanks is mediated by tracer particles,
which are injected at characteristic temperatures and rates
from heat baths at the two ends of the chain. 
For stochastic and Hamiltonian models of this type, we develop 
a theory that allows one to derive rigorously -- under
physically natural assumptions -- macroscopic equations for quantities  
related to heat transport, including mean energy profiles 
and tracer densities. Concrete examples are treated for illustration, 
and the validity of the Fourier Law in the present context is discussed. 

\end{abstract}
\thispagestyle{empty}
\tableofcontents


\section{Introduction}\label{intro}

Heat conduction in solids has been a subject of intensive study ever
since Fourier's pioneering work. An interesting issue 
is the derivation of macroscopic conduction laws from the microscopic
dynamics describing the solid. A genuinely realistic model of the solid
would involve considerations of 
quantum mechanics, radiation and other phenomena. 
In this paper, we address a simpler set of questions, viewing solids
that are effectively 1-dimensional as modeled by chains 
of classical Hamiltonian systems in which heat transport is mediated by 
tracer particles.  Coupling the two ends of the chain to unequal 
tracer-heat reservoirs and allowing the system to settle down to 
a nonequilibrium steady state, we study the distribution 
of energy, heat flux, and tracer flux  in this context.

We introduce in this paper a class of models that can be seen as
an abstraction of certain types of mechanical models.
These models  are simple enough to be amenable to analysis,
and complex enough to have fairly rich dynamics.
They have in common the following basic set of 
characteristics: Each model is made up of an array of identical cells 
that are linearly ordered. Energy is carried by two types of agents:
storage receptacles (called ``tanks") that are fixed in place, 
and tracer particles that move about. Direct energy exchange is 
permitted only between tracers and tanks. The two ends of 
the chain are coupled to infinite reservoirs that emit tracer particles 
at characteristic rates and characteristic temperatures;
they also absorb those tracers that reach them.
To allow for a broad range of examples, we do not specify the 
rules of interaction between tracers and tanks. All the rules
considered in this paper
have a Hamiltonian character, involving the kinetic energy
of tracers. Formally they may be stochastic or purely dynamical,
resulting in what we will refer to as {\it stochastic} and 
{\it Hamiltonian models}.

Via the models in this class, we seek to clarify the relation among 
several aspects of conduction, including the role of conservation laws, 
their relation to the dynamics within individual cells,  and the notion of 
``local temperature". We propose a simple recipe for deducing 
various macroscopic profiles from local rules (see Sect.~\ref{proposed}).
Our recipe is generic;  it does not depend on specific characteristics 
of the system. When the local rules are sufficiently simple, it produces 
explicit formulas that depend on exactly 4 parameters: the temperatures 
and rates of tracer injection at the left and right ends of the chain.

For demonstration purposes, we carry out this proposed program for a few
examples. Our main stochastic example, dubbed the 
``random-halves model", is particularly simple: A clock rings with
rate proportional to  $\sqrt x$ where $x$ is the (kinetic) energy 
of the tracer; at the clock, energy exchange between tracer and tank 
takes place; and  the rule of exchange consists simply of pooling 
the two energies together and randomly dividing -- in an unbiased way 
-- the total energy into two parts. Our main Hamiltonian example is 
a variant of the model studied in \cite{Larralde2002,Larralde2003}. 
Here the role of the ``tank" is played by a rotating disk nailed down at 
its center, and stored energy is $\omega^2$ where $\omega$ is
the angular velocity of the disk. Explicit formulas for the profiles
in question are correctly predicted in all examples.

In terms of methodology, this paper has a theory part and a simulations
part. The theory part is rigorous in the sense that all points that are 
not proven are isolated and stated explicitly as ``assumptions"
(see the next paragraph).  It also serves to elucidate 
the relation between various
concepts regardless of the extent to which 
the assumed properties hold. Simulations are used 
to verify these properties for the models considered. 

Our main assumption is in the direction of local thermodynamic equilibrium. 
For our stochastic models, a proof of this property seems within reach 
though technically involved (see {\it e.g.} \cite{Demasi1991,Spohn1991,Kipnis1999});
no known techniques are available for Hamiltonian systems.  
Extra assumptions pertaining to ergodicity and mixing issues are needed for
our Hamiltonian models.  It is easy to ``improve ergodicity" via
model design, harder to mathematically eliminate the possibility of all (small) 
invariant regions. In the absence of perfect mixing within cells, actual profiles 
show small deviations from those predicted for the ideal case.

\medskip
{\it In summary, we introduce in this paper a relatively tractable class
of models that can be seen as paradigms for heat conduction,
and put forth a program which -- under natural assumptions -- takes one 
from the microscopic dynamics of a system to its
phenomenological laws of conduction.}


\section{Main Ideas}\label{gsmi}

\subsection{General setup}\label{gsmodm}

The models considered in this paper -- both stochastic and Hamiltonian --
have in common a basic set of characteristics which we now describe.

\medskip
There is a finite, linearly ordered collection of
{\it sites} or {\it cells} labeled $1,2,\dots, N$. In isolation, \ie, when
the chain is not in contact with any external heat source, the system 
is driven by the interaction between two distinct types of energy-carrying 
objects:
\SI{~$\bullet$}
\begin{itemize}
\item[$\bullet$] Objects of the first kind are {\it fixed in place}, and
there is exactly one at each site. These objects play the role of storage 
facility, and serve at the same time to mark the energy level
at fixed locations. For brevity 
and for lack of a better word, let us call them energy {\bf tanks}.  
Each tank holds a finite amount of energy
at any one point in time; it is not to be confused with an infinite
reservoir.  We will refer to the energy in the tank at site $i$
as the {\bf stored energy} at site $i$.
\item[$\bullet$] The second type of objects are moving particles 
called {\bf tracers}. Each tracer carries with it a finite amount of energy,
and moves from site to site. For definiteness, we assume that from 
site $i$, it can go only to sites $i\pm 1$.
\end{itemize}
With regard to {\bf microscopic dynamics}, the following is assumed:
When a tracer is at site $i$, it may interact -- possibly multiple times -- with
the tank at that site. In each interaction, the two energies are pooled 
together and redistributed, so that energy is conserved in each interaction. 
The times of interaction and manner of redistribution are
determined by the microscopic laws of the system, which depend
solely on conditions within that site.  These laws determine
also the exit times of the tracers and their next locations.
{\it A priori} there is no limit to how many tracers are allowed
at each site.  We stress that this tracer-tank interaction is the only type 
of interaction permitted:  the tanks at different sites can communicate
with each other only via the tracers, 
and the tracers do not ``see" each other directly.

All {\it stochastic} models considered in this paper are Markovian.
Typically in stochastic rules of interaction, energy exchanges occur
when exponential clocks ring, and energy is redistributed
according to probability distributions. In {\it Hamiltonian} models, 
tracers are usually embodied by real-life moving
particles, and energy exchanges usually involve some types of collisions.

\medskip
The two ends of the chain above are coupled to two {\bf heat baths},
which are infinite
reservoirs emitting tracers at characteristic temperature (and also
absorbing them).   It is sometimes convenient to think of them as located at sites
$0$ and $N+1$. The two baths inject tracers into the system according to
certain rules (to be described). Tracers at site $1$ or $N$ can exit the
system; when they do so, they are absorbed by the baths.
The actions of the two baths are assumed to be independent of each
other and independent of the state of the chain. The left bath is set 
at temperature $\tl $; the energies of the tracers it injects into the system 
are {\it iid} with a law depending on the model. These tracers are injected 
at exponential rates, with mean $\jl $. Similarly the bath on the right 
is set at temperature $\tr $ and injects tracers into the system at rate $\jr $.

\medskip
To allow for a broad spectrum of possibilities, we have deliberately 
left unspecified (i) the rules of interaction between tracers and tanks, and
(ii) the coupling to heat baths, {\it i.e.}, the energy distribution of the
injected tracers. (Readers who wish to see concrete examples
immediately can skip ahead with no difficulty  to  Sects.~\ref{rh} and
\ref{fam}, 
where two examples are presented.) We stress that once (i) and (ii)
are chosen, and the 4 parameters $\tl , \tr , \jl $ and $\jr $ are set,
then all is determined:
the system will evolve on its own, and there is to be no
other intervention of any kind.

\begin{remark} \rm
Our approach can be viewed as that of a grand-canonical ensemble, 
since we fix the rates at which tracers are injected into the system 
(which indirectly determine the density and energy flux at steady state).
An alternate setup would be one in which the density of tracers 
is given, with particles being replaced upon exit. In this alternate
setup, the 4 natural extensive variables would be the temperatures 
$\tl$ and $\tr$, the density of tracers (mean number of tracers per cell) 
and the mean energy flux. For definiteness, we will adhere to
our original formulation.
\end{remark}

We now introduce  the {\bf quantities of interest}. 
For fixed $N$, let $\mu_N$ denote the invariant measure corresponding to
the unique steady state of the $N$-chain (assuming there is a unique
steady state). 
The word ``mean" below refers to averages
with respect to $\mu_N$. The main quantities of interest in this paper are
\SI{\ $\bullet$}
\begin{itemize}
\item[$\bullet$] \ $s_i\ = $ mean stored energy at site $i$~;
\item[$\bullet$] \ $e_i\ = $ mean energy of individual tracers at site $i$~;
\item[$\bullet$] \ $k_i\ = $ mean number of tracers at site $i$~;
\item[$\bullet$] \ $E_i\ = $ mean total energy at site $i$, including
stored energy and the energies of all tracers present.
\end{itemize}
For simplicity, we will refer to $e_i$ as {\it tracer energy}
and $E_i$ as {\it total-cell energy}.

We are primarily interested in the {\bf profiles} of these quantities,
\ie, in the functions $i \mapsto s_i, e_i, k_i$ and $E_i$ as $N \to \infty$
 {\em with the temperatures and injection rates of the baths
held fixed}. More precisely,  we fix $\tl , \tr , \jl $ and $\jr $. Then spacing
$\{1,2,\dots,N\}$ evenly along the unit interval $[0,1]$ and letting
$N \to \infty$, the finite-volume profiles $i \mapsto s_i, e_i, k_i, E_i$
give rise to functions $\xi \mapsto s(\xi), e(\xi), k(\xi), E(\xi)$,
$\xi \in [0,1]$. It is these functions that we seek to
predict given the microscopic rules that define a system.


\subsection{Proposed program: from local rules to global profiles}\label{proposed}

We fix $N$, $\tl , \tr , \jl $ and $\jr $, and consider an $N$-chain with
these parameters.  To determine the profiles in Sect.~\ref{gsmodm},
we distinguish between the following two kinds of information:
\SI{\ (b)}
\begin{itemize}
\item[(a)] cell-to-cell traffic, and
\item[(b)] statistical information pertaining to the dynamics
within individual cells.
\end{itemize}
In (a), we regard the cells as black boxes, and observe only what
goes in and what comes out. Standard arguments balancing energy and tracer
fluxes give easily the mean energy and number of tracers transported from
site to site. While these numbers are  indicative of the internal states of
the cells (for example, high-energy tracers emerging from a cell suggests
higher temperatures inside), the profiles we seek depend on more intricate
relations than these numbers alone would tell us.

We turn therefore to (b). Our very na\"ive idea is to study a {\it single
cell}, and to bring to bear on chains of arbitrary length the information
 so obtained. We propose the following plan of action:
\SI{\ (iii)}
\begin{itemize}
\item[(i)] Consider a single cell plugged to two heat baths (one on its left, 
the other on its right), both of which are 
at temperature $T$ and have injection rate $\rate$, $T$ and $\rate$ being
arbitrary.
Finding the invariant measure $\mu^{\Tr }$ describing the state of
the cell in this equilibrium situation is,
in general, relatively simple compared to finding $\mu_N$.

\item[(ii)] Suppose the measure $\mu^{\Tr}$ has been found. We then
look at an $N$-chain with $\tl =\tr =T$ and $\jl=\jr =\rate$, and verify 
that the marginals at site $i$ of the invariant measure $\mu_N$
are equal to $\mu^{\Tr }$. (By the marginal at site $i$,
we refer to the measure obtained by integrating out all variables pertaining
to all sites $\neq i$.)

\item[(iii)] Once the family $\{\mu^{T,\j}\}$ is found and (ii) verified,  
we {\it assume} that the
structure common to the $\mu^{\Tr }$ passes to all marginals of 
$\mu_N$ as $N \to \infty$ even when $(\tl , \jl) \neq (\tr , \jr )$.
More precisely, for all $\xi \in (0,1)$, we assume that all limit
points of the marginals of $\mu_N$ at site $[\xi N]$ (where $[x]$
denotes the integer part of $x$) inherit, as $N \to \infty$,
the structure common to $\mu^{\Tr }$. 
\end{itemize}

We observe that (i)--(iii) alone are  inadequate for determining the
sought-after profiles, for they give no information on which $T$ and 
$\rate$ are relevant at any given site. The main point of this program
is that (a) and (b) {\it together} is sufficient for
uniquely determining the profiles in question.

\begin{remark}\rm
Our rationale for (iii) is as follows:
Fix an integer, $\ell$. As $N \to \infty$, the gradients of temperature and
injection rate on the $\ell$ sites centered at $[\xi N]$ tend to $0$, so that
the subsystem consisting of these $\ell$ sites resembles more and more
the situation in (ii). Though rather natural from the point of view of physics
\cite{Groot1962}, this argument does not constitute a proof.
Indeed our program
is in the direction of proving the existence of well defined Gibbs
measures and then assuming, when the system is taken out of equilibrium,
that thermodynamic equilibrium is attained locally; in particular,
local temperatures are well defined. The full force of \LTE/ is not needed
for our purposes; however. The assumption in (iii) 
pertains only to marginals at single sites.
\end{remark}

The rest of this paper is devoted to illustrating the program outlined 
above in concrete examples. 


\section{Stochastic Models}\label{stoch}

\subsection{The ``random-halves" model}\label{rh}

This is perhaps the simplest stochastic model of the general type 
described in Sect.~\ref{gsmodm}. The microscopic laws that govern 
the dynamics in each cell are as follows:
Let $\delta>0$ be a fixed number. Each tracer is equipped 
with two independent exponential clocks. Clock 1, which signals the
times of energy exchanges with the tanks, rings
at rate $\frac{1}{\delta} \sqrt x$ where $x$
is the (current) energy of the tracer. Clock 2, which signals
the times of site-to-site movements, rings at rate $\sqrt{x}$.
The stored energy at site $i$ is denoted by $y_i$.
In the description below, we assume
 the tracer is at site $i$. 
\begin{itemize}
\item[(i)] When Clock 1 rings, the energy carried by the tracer and
the stored energy at site $i$ are pooled together and split randomly.
That is to say, the tracer gets $p(x+y_i)$ units of energy and the tank gets
$(1-p)(x+y_i)$, where $p \in [0,1]$ is uniformly
distributed and independent of all other random variables.
\item[(ii)] When Clock 2 rings, the tracer leaves site $i$.
It jumps  with equal probability to sites $i \pm1$.
If $i=1$ or $N$, going to sites $0$ or $N+1$ means the tracer exits the system.
\end{itemize}
It remains to specify the coupling to the heat baths. Here it is natural
to assume that the energies of the emitted tracers are exponentially 
distributed with means $\tl$ and $\tr$.

\medskip
This completes the formal description of the model.

\begin{remark}\rm
The rates of the two clocks are to be understood as
follows: We assume the energy carried by the tracer is
purely kinetic, so that its speed is $\sqrt x$. We assume also that a tracer
travels, on average, a distance $\delta$ between successive interactions 
with the tank, and a distance $1$ before exiting each site. 
\end{remark}
\begin{remark}\rm
As we will show, the invariant measure does not depend on 
the value of $\delta$, which can be large or small. 
The size of $\delta$ does affect the  rate of convergence to 
equilibrium, however.
\end{remark}

\begin{remark}\rm
While the tracers do not ``see" each other in the sense
that there is no direct interaction, their evolutions cannot be decoupled.
The number of tracers present at a site varies with time. When two or 
more tracers are present, they interact with the tank whenever their 
clocks go off, thereby sharing information about their energies. 
A new tracer may enter at some random moment,
bringing its energy to the pool; just as randomly, a tracer leaves, 
taking with it the energy it happens to be carrying at that time.
\end{remark}


\subsection{Single-cell analysis}\label{ssa}

\subsubsection{Single cell in equilibrium with 2 identical heat baths}

We consider first the following special case of the model described in
Sect.~\ref{rh}: $N=1$, $\tl =\tr =T$, and $\jl =\jr =\j$.
Each state of the cell in this model is represented by  a point in
$$
\Omega \ = \ \bigcup_{k=0}^\infty  \Omega_k \qquad {\rm (disjoint \ union)}
$$
where $\Omega_k=\{(\{x_1, \dots, x_k\}, y): x_\l , y \in [0,\infty)\}$. Here
$\{x_1, \dots x_k\}$ is an {\it unordered} $k$-tuple representing
the energies of the $k$ tracers, $y$ denotes the stored energy, and
a point in $\Omega_k$ represents a state of the cell when exactly $k$ tracers
are present.

\begin{remark} \rm
We motivate our choice of $\Omega$.
During a time interval when there are exactly $k$ tracers
in the cell -- with no tracers entering or exiting -- it makes little
difference whether
we think of the tracers as {\it named}, and represent
the state of the cell by a point in $[0,\infty)^{k+1}$, or if we think of them
as
{\it indistinguishable}, and represent the state by a point in $\Omega_k$.
With tracers entering and exiting, however, thinking of tracers as named
will require that all exiting tracers return later, otherwise
the system is transient and has no invariant measure.
Since any rule that assigns to each departing tracer a new tracer to carry its
name is
necessarily artificial, and for present purposes exact identities of tracers
play no role, we have opted to regard the tracers as indistinguishable.
\end{remark}

We clarify the relationship between $[0,\infty)^{k+1}$ and $\Omega_k$
and set some notation: Let $\pi_k: [0,\infty)^{k+1}$ $\to \Omega_k$ 
be the map $\pi_k(x_1, \dots, x_k, y) =(\{x_1, \dots x_k\}, y)$, \ie, $\pi_k$ 
is the $(k!)$-to-$1$ map that forgets the order in the ordered $k$-tuple
$(x_1, \dots, x_k)$. For a measure $\tilde \mu$ on $[0,\infty)^{k+1}$ 
that is symmetric with respect to the $x_\l $ coordinates, if
$\mu=(\pi_k)_*\tilde \mu$, and $\tilde \sigma$ and $\sigma$ 
are the densities of $\tilde \mu$ and $\mu$ respectively, then
$\tilde \sigma$ and $\sigma$  are related by
$$
\sigma(\{x_1, \dots, x_k\}, y) \ = \ k! \ \tilde \sigma(x_1, \dots, x_k, y) ~.
$$
We also write  $d\{x_1,\dots, x_k\} dy = 
(\pi_k)_*(dx_1 \dots dx_k dy)$, and use $I$ to denote the 
characteristic function.

\begin{proposition}\label{p1} The model in
Sect.~\ref{rh} with $N=1$, $\tl =\tr =T$, and $\jl =\jr =\j$ has a unique
  invariant probability measure $\mu=\mu^{\Tr}$ on $\Omega$. This measure 
  has the following properties:
\SI{\ $\bullet$}
\begin{itemize}
\item[$\bullet$]the number of tracers present is a Poisson random variable
with mean $\kappa\equiv 2\rate\sqrt {\pi/ T}$, \ie,
\begin{equation}\label{(a)}
\mu(\Omega_k) \ = \ \frac{\kappa^k}{k!} e^{-\kappa}~, 
\qquad k=0,1,2,\ldots ~;
\end{equation}
\item[$\bullet$]the conditional density of $\mu$ 
on $\Omega_k$ is $c_k \sigma_k d\{x_1, \dots, x_k\} dy$
where
\begin{equation}\label{(b)}
\sigma_k(\{x_1, \dots, x_k\}, y) \ = \ I_{\{x_1, \dots, x_k, y \geq 0\}} \
\frac{1}{\sqrt{x_1 \cdot\ldots\cdot x_k}} \ e^{-\beta (x_1+\dots + x_k +y)}~;
\end{equation}
here $\beta ={1}/{T}$, and
$c_k=\beta \,k! \left(\beta /\pi\right)^{k/2}$ is the normalizing
constant.
\end{itemize}
\end{proposition}

\noindent {\bf Proof:} Uniqueness is straightforward, since one can go
from a neighborhood of any point in $\Omega$ to a neighborhood of any other
point via positive measure sets of sample paths. We focus on checking
the invariance of $\mu$ as defined above.

\medskip
For $z, z' \in \Omega$, let $P^h(dz'|z)$
denote the transition probabilities for time $h\ge0$ starting from $z$.
We fix a small cube $A \subset \Omega_{\bar k}$ for some $\bar k$, 
and seek to prove that
$$
\frac{d}{dh} \ 
\left .\int \left( \int I_{A}(z') P^h(dz'|z) \right) \mu(dz)\right |_{h=0} \
= \ 0~.
$$
On the time interval $(0,h)$, the following three types of events may
occur:

\medskip
Event $E_1$: Entrance of a new tracer

\smallskip
Event $E_2$: Exit of a tracer from the cell

\smallskip
Event $E_3$: Exchange of energy between a tracer and the tank

\medskip
\noindent We claim that with initial distribution $\mu$, the probability of 
more than one of these events occurring before time $h$ is $\oo(h)$ 
as $h \to 0$. This assertion applies to events both of the same type 
and of distinct types. It follows primarily from the fact that the 
these events are independent and occur at exponential rates. 
Of relevance also are the exponential tails of $\sigma_k$ 
and the Poisson distribution of 
$p_k:= \mu(\Omega_k)$ in the definition of $\mu$.
To illustrate the arguments involved, we will verify at the end 
of the proof that the probability of two or more tracers exiting 
on the time interval $(0,h)$ is $\oo(h)$,
but let us accept the above assertion for now and go on with the main
argument. 

Starting from the initial distribution $\mu$, we let $\Pr(E_i, A)$ denote 
the probability that $E_i$ occurs before time $h$ resulting in
 a state in $A$, and let $\Pr(E^c_1 \cap E_2^c \cap E^c_3, A)$
denote the probability of starting from a state in $A$ and having
 none of the $E_i$ occur before time $h$. We will prove
\begin{equation}\label{e:33}
\Pr(E_1,A) + \Pr(E_2,A) + \Pr(E_3,A) + \Pr(E^c_1 \cap E_2^c \cap E^c_3, A)
-  \mu(A) = \oo(h) \mu(A)~.
\end{equation}
Notice that $A$ can be represented as the union of disjoint sets 
$\cup_i A_i$ where each $A_i$ is of the form
$$A_\varepsilon(\bar z) = \{ (\{x_1, \dots, x_{\bar k}\}, y): x_\l  \in
[\bar x_\l , \bar x_\l +\varepsilon],\l=1,\dots,\bar k, y \in [\bar y, \bar
y+\varepsilon]\}$$
for some $\bar z=(\{\bar x_1, \dots, \bar x_{\bar k}\}, \bar y) \in \Omega$,
$\varepsilon \ll 1$, and with the intervals 
$[\bar x_\ell, \bar x_\ell +\varepsilon]$ pairwise disjoint
for $\ell=1,2, \dots, \bar k$. To prove \eref{e:33} for $A$, it suffices
to prove it for each $A_i$ provided $\oo(h)$ in \eref{e:33} is uniformly
small for all $i$.

\medskip
We consider from here on $A=A_\varepsilon(\bar z)$ with
the properties above.
Let $\sigma$ denote the density of $\mu$.
With $\varepsilon$ sufficiently small, we have 
$\mu(A) \approx \sigma(\bar z)
\varepsilon^{\bar k+1} = p_{\bar k} c_{\bar k}
\sigma_{\bar k}(\bar z)\varepsilon^{\bar k+1}$
where $c_k$ and $\sigma_k$ are as in the proposition.
The other terms in \eref{e:33} are estimated as follows:

\bigskip
\noindent $\Pr(E_1,A)$: \ $E_1$ is in fact
the union of $2\bar k$ subevents, corresponding to a new tracer coming
 from the left or right bath and the $\bar k$ approximate values of
energy of the new tracer. For definiteness, we assume the new tracer
arrives from the left bath, and has energy in $[\bar x_1, \bar
x_1+\varepsilon]$.
That is to say, the initial state of the cell is described by
$$
B = \{(\{x_2, \dots, x_{\bar k}\}, y): x_\l  \in [\bar x_\l , \bar x_\l
+\varepsilon],
y \in [\bar y, \bar y+\varepsilon]\} \ \subset \ \Omega_{\bar k-1}~.
$$
The contribution to $\Pr(E_1,A)$ of this subevent is
\begin{eqnarray*}
\mu(B)  \ h\j  \ \int_{\bar x_1}^{\bar x_1+\varepsilon} \beta e^{-\beta x} dx
& \approx & h\j  \ \mu(B) \ e^{-\beta \bar x_1} \ \beta \varepsilon\\
& \approx & h\j  \ p_{\bar k-1} c_{\bar k-1}  \ \sigma_{\bar k}(\bar z) \
\sqrt{\bar x_1} \
\beta \varepsilon^{\bar k+1}~.
\end{eqnarray*}
Here, $h\j $ is the probability that a tracer is injected, and the integral
above
is the probability that the injected tracer lies in the specified range.
Summing over  all $2\bar k$ subevents, we obtain
\begin{equation} \Pr(E_1,A) \ \approx \ 
2h\j \beta \ (\sum_{\l=1}^{\bar k} \sqrt{\bar x_\l }) \
p_{\bar k-1} c_{\bar k-1}  \
\sigma_{\bar k} (\bar z) \ \varepsilon^{\bar k+1} ~.\label{e:1}
\end{equation}

\medskip
\noindent $\Pr(E_2,A)$: \  In order to result in a state in $A$, the 
initial state must be in
$$
C_1 = \{(\{x_1, \dots, x_{\bar k}, x\}, y): x_\l  \in [\bar x_\l , \bar x_\l
+\varepsilon], x \in [0,\infty),
y \in [\bar y, \bar y+\varepsilon]\} \ \subset \ \Omega_{\bar k+1}~.
$$
We assume here that the tracer with energy $x$ exits between time 
$0$ and time $h$. This gives 
\begin{eqnarray}
\Pr(E_2,A) & \approx  &  p_{\bar k+1} \ c_{\bar k+1} \ \sigma_{\bar k}(\bar z) \ \varepsilon^{\bar k+1}
\ \int_0^\infty \min\{h \sqrt {x}, 1\} \ \frac{1}{\sqrt x} e^{-\beta x} \ dx\nonumber\\
& =  & p_{\bar k+1} \ c_{\bar k+1} \ \sigma_{\bar k}(\bar z)
 \ \varepsilon^{\bar k+1} \ \beta ^{-1} (h + \oo(h))~.\label{e:1a}
\end{eqnarray}

\noindent $\Pr(E_3,A)$: \
For definiteness, we assume it is the tracer with energy near $ \bar x_1$
that is the product of the interaction with the tank. To arrive in a
state in $A_\varepsilon$, 
 one must start from
$$
D = \{ (\{x_1, \dots, x_{\bar k}\}, y) : x_\l  \in [\bar x_\l , \bar x_\l
+\varepsilon]
\ {\rm for} \ \ell \geq 2, x_1 + y \in [\bar x_1+\bar y, \bar x_1+\bar y +
2\varepsilon]\}~.
$$
A simple integration using the rule of interaction in Sect.~\ref{rh}
gives 
\begin{equation}\label{e:1b}
\Pr(E_3,A) \ \approx \ h   \frac{\sqrt{\bar x_1}}{\delta} \ p_{\bar k} c_{\bar k} \sigma_{\bar k}(\bar z)
\varepsilon^{\bar k+1}~.
\end{equation}

\medskip
\noindent $\Pr(E^c_1 \cap E_2^c \cap E^c_3, A)$: \ 
We first note that starting from $A$,
 the probability of the tracer with energy $\approx \bar x_1$ exiting is
$$
p_{\bar k} c_{\bar k} \sigma_{\bar k}(\bar z) \sqrt {\bar x_1} e^{\beta \bar x_1}
\varepsilon^{\bar k} \int_{\bar x_1}^{\bar x_1+\varepsilon}
h \sqrt x \frac{1}{\sqrt x} e^{-\beta x} dx
\ = \ h \sqrt {\bar x_1} \ p_{\bar k} c_{\bar k}  \sigma_{\bar k}(\bar z)
\ \varepsilon^{\bar k+1}~;
$$
\noindent the probability of the tracer with energy $\approx \bar x_1$
interacting with the tank is
$$
\ h \frac{\sqrt {\bar x_1}}{\delta} \ p_{\bar k} c_{\bar k}  \sigma_{\bar k}(\bar
z)
\ \varepsilon^{\bar k+1}~;
$$
and the probability of a new tracer entering the cell 
from the left (resp.~right)
bath is $h\j \ \mu(A_\varepsilon)$. Thus
\begin{eqnarray}\label{e:4}
\Pr(E^c_1 \cap E_2^c \cap E^c_3, A) & \approx &
 p_{\bar k} c_{\bar k} \sigma_{\bar k}(\bar z) \  \varepsilon^{\bar k+1} \cdot
\Pi_\l (1-h \sqrt{\bar x_\l }) \cdot (1-h\j )^2 \cdot 
\Pi_\l (1- {h}\sqrt{\bar x_\l }/\delta)\nonumber\\
& \approx & p_{\bar k}  c_{\bar k} \sigma_{\bar k}(\bar z) \  \varepsilon^{\bar
k+1} \cdot
 \left(1 - h \left(\Sigma_{\l=1}^{\bar k} \sqrt{\bar x_\l } + 2\rho +
 \frac{1}{\delta} \Sigma_{\l=1}^{\bar k} \sqrt{\bar x_\l }
\right) \right)\kern-0.3em.
\end{eqnarray}
Summing Eqs.~\eref{e:1}--\eref{e:4}, we obtain \eref{e:33} provided
$$
2\rate \beta  p_{\bar k-1} c_{\bar k-1}  \ = \ p_{\bar k} c_{\bar k}
\qquad {\rm and} \qquad
2\rate p_{\bar k} c_{\bar k}  \ = \ p_{\bar k+1} c_{\bar k+1} T~.
$$
Note that these two equations represent the same relation for different $k$.
We write this relation as
\begin{equa}[2][e:99]
\frac{c_k p_k}{c_{k+1}p_{k+1}}&=\frac{T}{2\rho}~,\\
\end{equa}
and verify that it is compatible with assertion \eref{(a)}: Since
$$
c_k \ \frac{1}{k!} \ \left(\Pi_{\l=1}^k \int \frac{1}{\sqrt{x_\l }}e^{-\beta
x_\l }dx_\l  \right)
\ \int e^{-\beta y}dy \ = \ 1~,
$$
and $\int_0^\infty x^{-1/2} e^{- \beta x}dx = \sqrt{\pi T }$,
we have
\begin{equa}[2]
p_{k+1} \ = \ \frac{2\rho}{T} \ \frac{c_k}{c_{k+1}} \ p_k
\ = \ \frac{2\rho}{T} \ \left(\frac{1}{k+1} \int_0^\infty \frac{1}{\sqrt x}
e^{-\beta x}dx
\right)  \ p_k \ = \ \frac{1}{k+1} \ \frac{2\sqrt{\pi} \rho}{\sqrt T} \ p_k~.
\end{equa}

\bigskip
To complete the proof, we estimate the probability of two or more
tracers exiting before time $h$. For  $n= 2,3, \dots$, let $E_{2,n}$ 
be the event that the initial state is in
$$
C_n = \{(\{x_1, \dots, x_{\bar k}, x^{(1)}, \dots, x^{(n)}\}, y):
x_\l  \in [\bar x_\l , \bar x_\l +\varepsilon], x^{(\l')} \in [0,\infty),
y \in [\bar y, \bar y+\varepsilon]\}~,
$$
and during the time interval $(0,h)$, all $n$ of the tracers carrying
energies $x^{(\l')}, \l'=1,2, \dots, n$, exit the system.  The probability of
$\cup_{n \geq 2} E_{2,n}$ is
$$
\sum_{n \geq 2} p_{\bar k+n} c_{\bar k+n} \sigma_{\bar k}(\bar z)
\varepsilon^{\bar k+1} \ (\OO(h))^n~,
$$
which, from Eq.~\eref{e:99}, is bounded by
$$
p_{\bar k} c_{\bar k} \sigma_{\bar k} (\bar z) \varepsilon^{\bar k+1} \
\sum_{n \geq 2} \left(\frac{2\rho}{T}\right)^n (\OO(h))^n \ = \
p_{\bar k} c_{\bar k} \sigma_{\bar k} (\bar z) \varepsilon^{\bar k+1} \ \oo(h)~.
$$
 \hfill $\square$

\begin{remark}\label{r:35}\rm In the setting of Proposition~\ref{p1}, since the cell is
in equilibrium with the two heat baths, it is obvious that it ejects, on 
average, $2\j$ tracers per unit time, and the energies of the tracers 
ejected have mean $T$. We observe that the cell in fact reciprocates
the action of the bath more strongly than this: the {\it distribution} of 
the energies of the ejected tracers is also exponential. To see
this,  fix $k$ and consider one tracer at a time. The probability 
of the tracer exiting with energy $>u$ is
$$
\sim \ \int_u^\infty \sqrt x \cdot \frac{1}{\sqrt x} e^{-\beta x} dx \ = \ 
\beta ^{-1}e^{-\beta u}~.
$$
\end{remark}

\subsubsection{Chain of $N$ cells in equilibrium with $2$ identical
  heat baths}

We treat next the case of arbitrary $N$. That is to say, the system is
as defined in Sect.~\ref{rh}, but with $\tl =\tr =T$ and $\jl =\jr =\j$.
Let $\mu$ be as in Proposition \ref{p1}.

\begin{proposition}\label{p2}
The $N$-fold product $\mu \times \dots \times \mu$ is invariant.
\end{proposition}

\begin{remark}\rm
That the invariant measure is a product tells us that at
steady state, there are no spatial correlations. We do not find this
to be entirely 
obvious on the intuitive level: one might think that above-average 
energy levels on the left half of the chain may cause the right
half to be below
average; that is evidently not the case. This result should not be 
confused with the absence of {\it space-time} correlations.
\end{remark}

\noindent {\bf Proof:} \ Proceeding as before, we consider a small time
interval
$(0,h)$, and treat separately the individual events that may occur during this
period.
One of the new events (not relevant in the case of a single cell) is the
jumping
of a tracer from site $i \pm1$ to site $i$. We fix a phase point
$$
\bar z \ = \ (\bar z^{(1)}, \dots, \bar z^{(N)}) \ = \
(\{\bar x^{(1)}_1, \dots, \bar x^{(1)}_{k_1}\}, y^{(1)}; \ \dots; \
\{\bar x^{(N)}_1, \dots, \bar x^{(N)}_{k_N}\}, y^{(N)})~,
$$
and let $A = \Pi_{i=1}^N A^{(i)} \subset \Omega^N$
where $A^{(i)}=A_\varepsilon(\bar z^{(i)})$ is as in Proposition \ref{p1}. 
Let $\sigma^{(i)}=\sigma_{k_i}(\bar z^{(i)})$.
For definiteness, we fix also an integer
$n, 1<n<N$, and assume that at time $0$, 
the state of the chain is as follows:

\SI{(ii)}
\begin{itemize}
\item[(i)]at site $n+1$, there are $k_{n+1}+1$ tracers the energies of which
lie in disjoint intervals
$$[\bar x^{(n+1)}_1, \bar x^{(n+1)}_1+\varepsilon], \ \dots, \
[\bar x^{(n+1)}_{k_{n+1}}, \bar x^{(n+1)}_{k_{n+1}}+\varepsilon] \
\quad {\rm and} \quad [\bar x^{(n)}_1, \bar x^{(n)}_1+\varepsilon]~,
$$
\item[(ii)]at site $n$, there are $k_n-1$ tracers whose energies lie in
$$[\bar x^{(n)}_2, \bar x^{(n)}_2+\varepsilon], \ \dots, \
[\bar x^{(n)}_{k_{n}}, \bar x^{(n)}_{k_{n}}+\varepsilon]~.$$
\end{itemize}
\noindent The probability of this event occurring {\it and} the tracer
with energy $\approx \bar x^{(n)}_1$ jumping from site $n+1$ into site
$n$ is then given by
$\frac{1}{2} I \cdot I\kern -0.3emI \cdot I\kern -0.3emI\kern -0.3emI$ where
\begin{eqnarray*}
I & = & \Pi_{i \neq n, n+1} \ (p_{k_i}c_{k_i} \sigma^{(i)}
\varepsilon^{k_i+1})\\
I\kern -0.3emI & = & p_{k_n-1} c_{k_n-1} \sigma^{(n)} \sqrt{\bar x^{(n)}_1}
e^{\beta \bar x^{(n)}_1}
\varepsilon^{k_n}\\
I\kern -0.3emI\kern -0.3emI & = & p_{k_{n+1}+1} c_{k_{n+1}+1} \sigma^{(n+1)}
\varepsilon^{k_{n+1}+1}
\ \int_{\bar x^{(n)}_1}^{\bar x^{(n)}_1+\varepsilon} h 
\sqrt x \frac{1}{\sqrt x} e^{-\beta x} dx~.
\end{eqnarray*}
This product can be written as
$$
\frac{h}{2} \ \left(\Pi_{i=1}^N p_{k_i}c_{k_i} \sigma^{(i)} \varepsilon^{k_i+1}
\right)
\cdot \frac{p_{k_n-1}}{p_{k_n}} \frac{c_{k_n-1}}{c_{k_n}}
\cdot \frac{p_{k_{n+1}+1}}{p_{k_{n+1}}} \frac{c_{k_{n+1}+1}}{c_{k_{n+1}}}
 \cdot \sqrt{\bar x^{(n)}_1}~,
$$
which, by Eq.~\eref{e:99}, is equal to 
\begin{equation}\label{e:39}
\frac{h}{2} \ \left(\Pi_{i=1}^N p_{k_i}c_{k_i} \sigma^{(i)} \varepsilon^{k_i+1}
\right)  \cdot \sqrt{\bar x^{(n)}_1}~.
\end{equation}

There are many terms of this kind that contribute to
$\int I_{A}(z') P^h(dz'|z) \mu(dz)$,
two for $\bar x^{(n)}_\ell$ for each pair $(n,\ell)$.
We claim that the system has detailed balance, {\it i.e.}, the 
term associated with the scenario above is balanced
by the probability of starting from a state in $A$ and having the tracer 
at site $n$ carrying energy $\approx \bar x^{(n)}_1$ jump to site $n+1$.
The probability of the latter is
$$
\frac{1}{2} (\Pi_{i=1}^N p_{k_i}c_{k_i} \sigma^{(i)} \varepsilon^{k_i+1}) \cdot
\sqrt{\bar x^{(n)}_1} e^{\beta \bar x^{(n)}_1} \frac{1}{\varepsilon} \cdot
\int_{\bar x^{(n)}_1}^{\bar x^{(n)}_1+\epsilon } h \sqrt x \frac{1}{\sqrt x}
e^{-\beta x}dx~,
$$
which balances exactly \eref{e:39} as claimed.

An argument combining the one above with that in Proposition \ref{p1} 
regarding the injection of new tracers holds at sites $1$ and $N$. 
 \hfill $\square$

\bigskip
Propositions \ref{p1} and \ref{p2} are steps (i) and (ii) in the proposed
scheme in Sect.~\ref{proposed}.


\subsection{Derivation of equations of macroscopic profiles}\label{profiles}

Having found a candidate family of equilibrium measures $\{\mu^{\Tr }\}$,
we now complete the rest of the program outlined in Sect.~\ref{proposed}.
The next step, according to this program, is to {\it assume} 
that for $N\gg 1$, the marginals of the invariant measure $\mu_N$ 
at site $i$ are approximately equal to
$\mu^{\Tr }$ for some $T=T_i$ and $\rate=\rate_i$. 
We identify those parts of our proposed theory that are not proved
in this paper and state them precisely as ``Assumptions".

\medskip
\begin{changemargin}{1cm}{1cm} 
{\bf Assumption 1.} \ {\it Given $\tl, \tr>0$,  $ \jl, \jr \geq 0$, and $N \in
{\mathbb Z}^+$, the $N$-chain defined in Sect.~\ref{rh}
with these parameters has  an invariant 
probability measure $\mu_N$.}
    \end{changemargin} 

\medskip
We do not believe this existence result is hard to prove but prefer not to 
depart from the main line of reasoning in this paper.
Once existence is established, uniqueness (or ergodicity) follows easily 
since any
invariant measure clearly has strictly positive density everywhere.
A proof of the statement in Assumption 2 below is more challenging.
For $\xi \in (0,1)$, let $\mu_{N, [\xi N]}$ denote the marginal of $\mu_N$
at the site $[\xi N]$. 

\medskip
\begin{changemargin}{1cm}{1cm} 
{\bf Assumption 2.} \  {\it For every  $\xi  \in (0,1)$, every limit
point as $N \to \infty$ of $\mu_{N, [\xi N]}$ is a member of the family 
$\{\mu^{\Tr}, \ T > 0,\rate \geq 0\}$.}
\end{changemargin}

\medskip
In Sect.~\ref{gsmodm}, we introduced four quantities of interest.
There is one that was somewhat ambiguously defined, namely $e_i$. 
Its precise meaning is as follows:
$e_i := \sum_{k=1}^\infty p_{i,k} e_{i,k}$ where $p_{i,k}$ is the probability
that the number of tracers at site $i$ is equal to $k$ and $e_{i,k}$
is  $\frac{1}{k}$ of the mean total tracer energy
conditioned on the number of tracers present being equal to $k$.

Theorem~\ref{thm1} is about the profiles of certain observables as $N \to
\infty$. We refer the reader to the end of Sect.~\ref{gsmodm} for the precise meaning
of the word ``profile" in the theorem.

\begin{theorem}\label{thm1} The following hold for the ``random-halves" 
model defined in Sect.~\ref{rh} with arbitrary $\tl, \tr, \jl, \jr$. 
Under Assumptions 1 and 2 above:
\SI{\ $\bullet$}
\begin{itemize}
\item[$\bullet$] the profile for the mean number of jumps  out of a site 
per unit time is
$$
j(\xi ) \ = \ 2\bigl( \jl   + (\jr -\jl )\xi \bigr)~.
$$
\item[$\bullet$] the profile for the mean total energy transported out of a
site
per unit time  is
$$
Q(\xi) \ = \ 2\bigl(\jl \tl  + (\jr \tr -\jl \tl )\xi\bigr)~;
$$
\item[$\bullet$] the profile for the mean stored energy at a site  is
$$
s(\xi ) \ = \ \frac{Q(\xi)}{j(\xi)} \ = \ \frac{\jl \tl  + (\jr \tr -\jl \tl )\xi }{\jl +(\jr -\jl )\xi }~;
$$
in the case $\jl =\jr $, this simplifies to
$s(\xi ) =  \tl  + (\tr -\tl )\xi~; $
\item[$\bullet$] the profile for mean tracer energy is $e(\xi )=\frac{1}{2}
s(\xi )~;$
\item[$\bullet$] the profile for mean number of tracers is
$$
\kappa(\xi ) \ = \ \sqrt{\frac{\pi}{s(\xi )}} \ j(\xi ) ~;
$$
\item[$\bullet$] the profile for mean total-cell energy is
$$
E(\xi )\ = \ s(\xi ) + \kappa(\xi )e(\xi ) \ = \ s(\xi )  + \frac{1}{2}
\sqrt{\pi s(\xi )}\  j(\xi )~.
$$
\end{itemize}
\end{theorem}

\noindent {\bf Proof of Theorem~\ref{thm1} :} We divide the proof into
the following three steps:

\medskip
\noindent {\it I. Information on single cells:} Items (i)--(iv) are strictly in
the domain of {\it internal cell dynamics}. The setting is that of 
Proposition \ref{p1}, and the results below are deduced (in straightforward
computations)  from the invariant density given by that proposition.
The parameters are, as usual, $T$ and $\rate$. (Note that this means
 the rate at which tracers enter the site is $2\rate$.)
\SI{\ (iii)}
\begin{itemize}
\item[(i)]  stored energy has density $\beta e^{-\beta y}$ and mean $T$~;

\item[(ii)] tracer energy has density $\frac{\sqrt \beta}{\sqrt{\pi x}} e^{-\beta x}$
and mean $\frac{T}{2}$~;

\item[(iii)] mean number of tracers, $\kappa = 2 \sqrt{\frac{\pi}{T}} \j$~;

\item[(iv)] mean total-cell energy, $E=T\cdot(1+ \frac{\kappa}{2})$~.
\end{itemize}
Items (v) and (vi) are in preparation for the analysis of {\it cell-to-cell
traffic}\,\/:
\begin{itemize}
\item[(v)] mean number of jumps out of the cell per unit time, $j=2\j$~;
\item[(vi)] mean total energy transported out of the cell per unit time,
$Q = Tj$. 
\end{itemize}

\medskip
\noindent {\it II. Global phenomenological equations:}
Consider now a chain with $N$ cells with settings $\tl, \tr, \jl$ and $\jr$
at the two ends. The following results use only 
standard conservation laws together with the local rule that when
a tracer exits a cell, it has equal probability of going left and right.

\bigskip
\noindent (A) {\it Balancing tracer fluxes:} Let $j_i$ denote the number of
jumps
per unit time out of site $i$. Then
\begin{equ}[e:100]
j_i = 2 \left(\jl  + \frac{i}{N+1} \ (\jr  - \jl ) \right)~.
\end{equ}
\noindent {\it Proof:} Consider an (imaginary) partition between site $i$ and
site $i+1$.
We let  $-\Delta j_i$ denote the flux across this partition.
Then $\Delta j_i = \frac{1}{2} (j_{i+1}-j_i)$ for $i=0,1, \dots, N$
where $j_0$ and $j_{N+1}$ are defined to be $2\jl $ and $2\jr $ respectively.
For $i \neq 0,N$, the $\frac{1}{2}$ is there because only half of the tracers
out
of site $i+1$  jump left, and half of those out of site $i$ jump right.
 The fluxes across partitions
between all consecutive sites must be equal, or there would be a pile-up of
tracers
somewhere. This together with  $\sum_i \Delta j_i = 2(\jr -\jl )$ gives
the asserted formula.

\bigskip
\noindent (B) {\it Balancing energy fluxes:} Let $Q_i$ denote
the mean total energy transported out of site $i$ per unit time. Then
\begin{equ}
Q_i = 2 \left(\jl \tl + \frac{i}{N+1} \ (\jr\tr  - \jl \tl) \right)~.
\end{equ}
\noindent {\it Proof:} The argument is identical to that in (A), with
$\Delta Q_i = \frac{1}{2}(Q_{i+1}-Q_i)$ and $Q_0$ and $Q_{N+1}$
defined to be $2\jl\tl$ and $2\jr \tr$ respectively.

\bigskip
\noindent {\it III. Combining the results from I and II:}  Fix $\xi \in (0,1)$.
Passing to a subsequence if necessary,  we have, by Assumption 2, 
$\mu_{N, [\xi N]} \to \mu^{T(\xi), \rate(\xi)}$ for some $T(\xi)$ and $\rate(\xi)$.
We identify the two numbers $T(\xi)$ and $\rate(\xi)$ as follows:

Let $j^N(\xi)$ denote the mean number of jumps out of site $[\xi N]$ per
unit time in the $N$-chain. Then by (A) in part II, 
$j(\xi) :=\lim_{N \to \infty} j^N(\xi) = 2\bigl( \jl   + (\jr -\jl )\xi \bigr)$.
This is, therefore, the mean number of jumps out of a cell with invariant 
measure $\mu^{T(\xi), \rate(\xi)}$. Similarly, we deduce that the mean
total energy transported out of a cell with the same invariant measure
is $Q(\xi) :=\lim_{N \to \infty} Q^N(\xi) = 2\bigl(\jl \tl  + (\jr \tr -\jl \tl )\xi\bigr)$.
Appealing now to the information in part I, we deduce from (v) and (vi)
that $\rate(\xi)=\frac{1}{2}j(\xi)$ and $T(\xi)= Q(\xi)/j(\xi)$.

The rest of the profiles follow readily: we read off $s=T$ from (i), 
and deduce the relation between 
$s$ and $e$ by comparing (i) and (ii). The profiles for $\kappa$
and $E$ follow from (iii) and (iv) together with our knowledge
of $\j$ and $T$.
The proof of Theorem~\ref{thm1}  is complete.
\hfill  $\square$

\begin{remark}\rm It is instructive to see what Theorem~\ref{thm1}  says in
  the special case when $\jl=0$. Since no particles are injected at
  the left end, clearly, $\tl$ cannot matter. But since tracers do exit
  from the left, one expects an energy flux across the system.
Upon substituting $\jl=0$ into the formulas above, one gets
$$
j(\xi)=2\jr \xi~,\quad e(\xi)=\textstyle{\frac{1}{2}} s(\xi)=\textstyle{\frac{1}{2}}\tr~,\quad
 \kappa(\xi)= 2\jr\xi\sqrt{
{\pi}/{\tr}}~,
$$
and an energy flux of $-\frac{1}{2}Q'(\xi)=- \jr\tr$.
\end{remark}

\subsection{Simulations}\label{simulation}

Numerical simulations are used to validate Assumptions 1 and 2
 in Theorem~\ref{thm1} .

We mention here only those details of the simulations that differ 
from the theoretical study. Needless to say, we work with a finite 
number of sites, usually 20. 
Simulations start in a random initial state, and are first run for a period
of time to let the system reach its steady state. All times are in number 
of events ($E_1$, $E_2$, and $E_3$). Up to half the simulation 
time is used to reach stationarity; statistics are then gathered 
during the remaining simulation time. Since total simulation time is finite, 
we find it necessary to take measures to deal with tracers of 
exceptionally low energy: these tracers appear to remain in a cell 
indefinitely, skewing the statistics on the number of tracers. 
We opted to terminate events 
involving a single tracer at a single site
after about 0.001 of total simulation time. 
This was done about 10 times in the course of $10^9$ events.

\begin{figure}
\C{\psfig{file=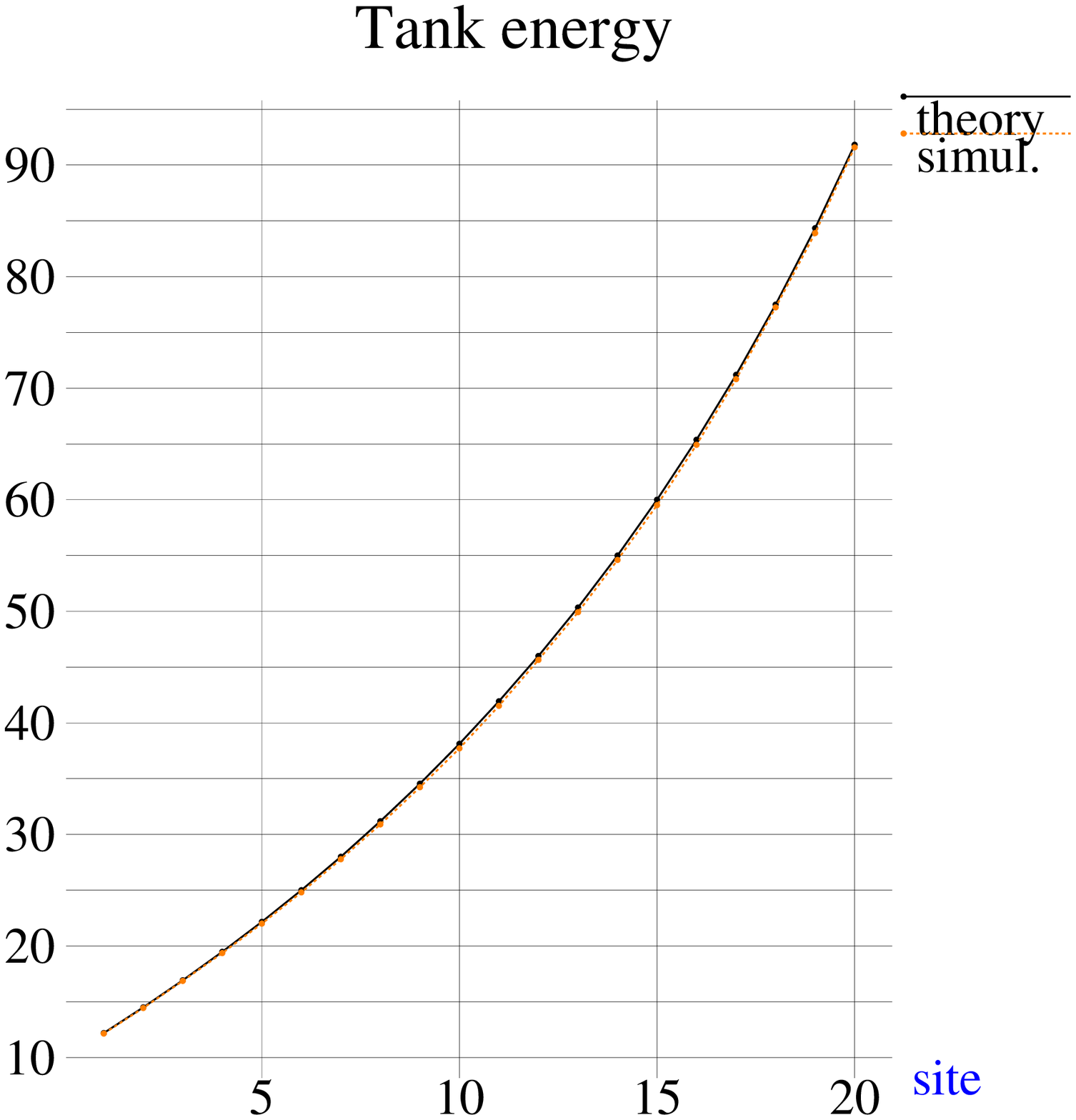,width=\x}\psfig{file=figs/x105.13.epsi,width=\x}}\C{\psfig{file=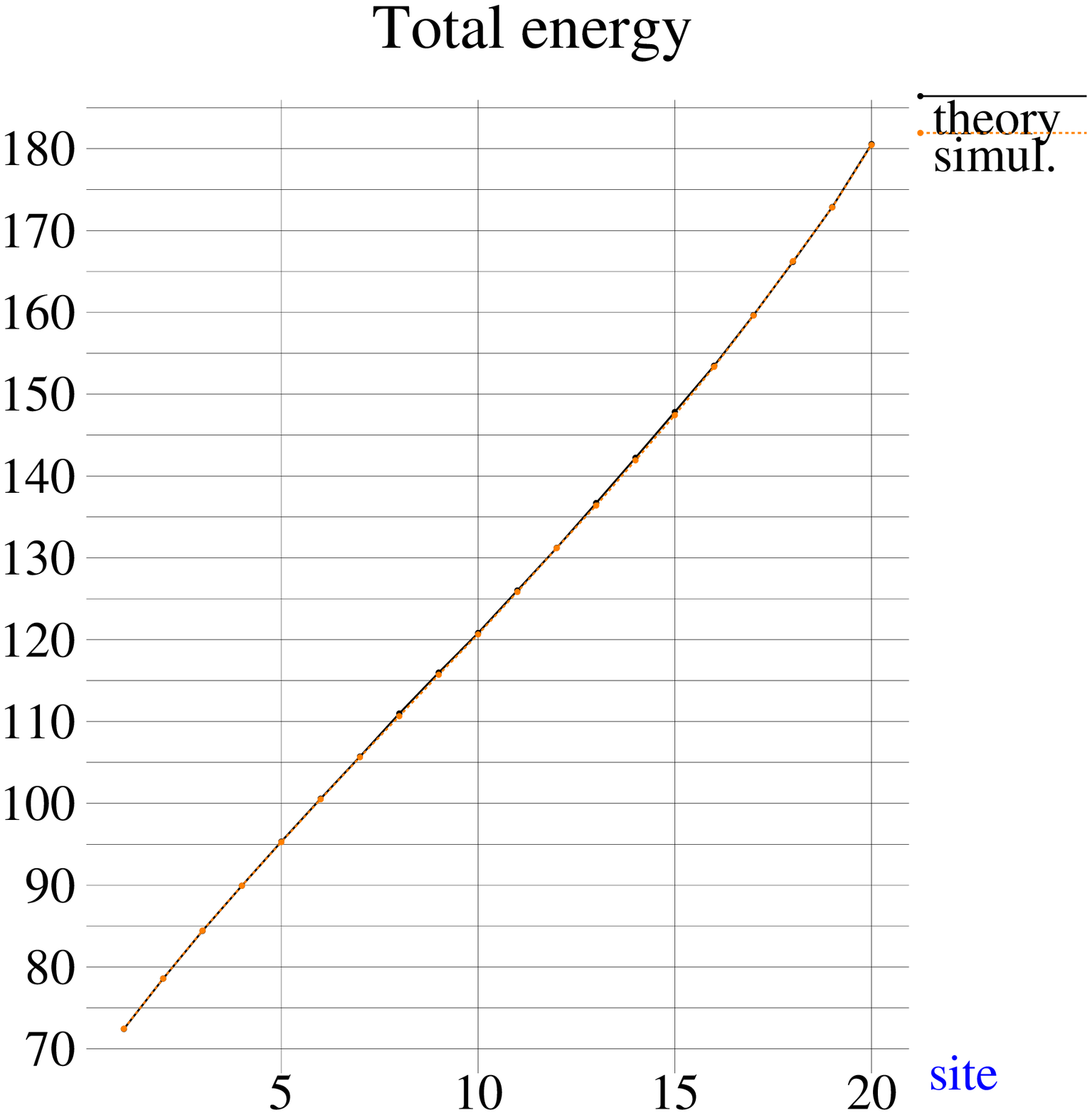,width=\x}}
\caption{Random-halves model with 20 sites, temperatures
$\tl=10$, $\tr=100$ and injection rates $\jl=10$, $\jr=5$.
Top left: Mean tank energies $s_i$.
Top right: Mean number of tracers $\kappa_i$.
Bottom: Mean total energy $E_i$.
Simulations in perfect agreement with predictions from theory.}\label{f:1051}
\end{figure}

Simulations are performed both to verify directly the properties of
the marginals at individual sites and to plot empirically the various
profiles of interest. Excellent agreement with predicted values is
observed in all runs. A sample of the results is shown in Fig.~\ref{f:1051}.

\subsection{Interpretation of results}\label{interpret}

We gather here our main observations, discuss their physical 
implications, and provide explanations for the reasons behind 
the derived formulas. The content of this subsection is valid not 
only for the random-halves model but for all of the examples 
studied in this paper. 

\bigskip
\noindent {\it 1. Linearity of profiles}

\medskip
We distinguish between the following 3 types of profiles:

\smallskip
\noindent a. {\it Transport of energy and tracers}:
$j(\xi)$ and $Q(\xi)$ are always linear
by simple conservation laws.

\smallskip
\noindent b. {\it Mean stored and (individual) tracer energies}:
$s(\xi)$ and $e(\xi)$ are linear if and only if there is 
no tracer flux across the system. (See item 4 below.)

\smallskip
\noindent c. {\it Tracer densities and total-cell energy}: 
$\kappa(\xi)$  is never linear (unless $\tl=\tr$ and $\jl=\jr$);
in addition to the obvious bias brought about by different injection
rates, tracers have a tendency to accumulate at the cold end
(see item 3 for elaboration). As a result, $E(\xi)$
is also never linear.

\bigskip
\noindent {\it 2. Heat flux and the Fourier Law}

\smallskip
Heat flux from left to right is given by $\Phi= \tl \jl -\tr \jr$.
Thinking of the temperature of the system as given by
 $T(\xi)$, Theorem~\ref{thm1}  says that thermal conductivity is constant 
 and proportional to 
$\tr-\tl$ if and only if there is no tracer flux across the system, {\it i.e.},
if and only if $\jl=\jr$.

\bigskip
\noindent {\it 3. Distribution of tracers along the chain} 

\smallskip
In the case $\jl =\jr $, more tracers are congregated at the cold end 
than at the hot. This is because the only way to balance the tracer 
equation is to have  the number of jumps out of a site be constant
along the chain.  Inside the cells, however,  tracers move more
slowly at the cold end, hence they jump less frequently, and
the only way to maintain the required number of jumps
is to have more tracers. When $\jl \neq \jr$, the
idea above continues to be valid, except that
one needs also to take into consideration the bias 
in favor of more tracers at the end where the injection rate is higher. 

\bigskip
\noindent {\it 4. Tracer flux and concavity of stored energy} \ 

\smallskip
One of the interesting facts that have emerged is that
$s(\xi)$ is linear if and only if $\jl =\jr $, and
when $\jl  \neq \jr $, their relative strength is
reflected in the concavity of $s(\xi)$. This may be a little perplexing
 at first
because no mechanism is built into the microscopic rules for the tanks to
recognize the directions of travel of the tracers with which they come into
contact.
The reason behind this phenomenon is, in fact, quite simple: If there is a
tracer flux
across the system, say from right to left, then the tank at site $i$
hears from site $i+1$ more frequently than it hears from site $i-1$
(because $j_{i+1}>j_{i-1}$). It therefore has a greater tendency to equilibrate
 with the energy level on the right than on the left, causing
 $s_i$ to be $> \frac{1}{2}(s_{i+1}+s_{i-1})$. Since this
happens at every site, a curvature for the profile of $s_i$ is created.
The reader should further note that tracer flux and heat flux go
in opposite directions if $(\rho_\L-\rho_\R)\cdot(\rho_\L T_\L-\rho_\R
T_\R)<0$.

\bigskip
\noindent {\it 5. Individual cells mimicking heat baths} \ 

\smallskip
The cells in our models are clearly not infinite heat reservoirs, yet
for large $N$, they acquire some of the 
characteristics of the heat baths with which they are
in contact. More precisely, the $i$th cell injects each of its two neighbors 
with $\j_i$ tracers per unit time. These tracers, which have mean energy 
$T_i$, are distributed according to 
a law of the same type as that with which tracers are emitted
 from the baths  
(exponential in the case of the random halves model);
see Remark~\ref{r:35}. Unlike the conditions at the two ends, however, 
the numbers $\j_i$ and $T_i$ are {\it self-selected}.


\subsection{A second example}\label{otherexamples}

We consider here a model of the same type as the ``random-halves"
model but with {\it different microscopic rules}. The purpose of this exercise
is to highlight the role played by these rules and to make transparent
which part of our scheme is generic.

\medskip
The rules for energy exchange in this model simulate a Hamiltonian model
in which both tracers and tanks have one degree of freedom.
 Write $x=v^2$ and $y=\omega^2$,
$v, \omega \in (-\infty, \infty)$, and think of energy as uniformly distributed
on the circle $\{(v,\omega): v^2+\omega^2=c\}$, so that when the tracer
interacts with stored energy, the redistribution  is such that a random point
on this circle is chosen with weight $|v|$ (this is the measure induced on
a cross-section by the invariant measure of the flow). That is to say,
if $(x,y)$ are the
stored energy and tracer energy before an interaction, and $(x',y')$
afterwards, then for $a \in [0, x+y]$,
\begin{equ}[e:exchange]
P\{y'>a\} \ = \ P\{|\omega'|> \sqrt a\} \ = \ 1 -
\frac{\int_0^{\sqrt a} v \sqrt{1+ \left(\frac{dv}{d\omega}\right)^2} d\omega}
{\int_0^{\sqrt{x+y}} v \sqrt{1+ \left(\frac{dv}{d\omega}\right)^2} d\omega}
\ = \ 1 - \frac{\sqrt a}{\sqrt{x+y}}~,
\end{equ}
or, equivalently, the density of $y'$ is
$$\frac{1}{2} \frac{1}{\sqrt{y'}} \frac{1}{\sqrt{x+y}}
 \qquad {\rm for} \quad y' \in [0, x+y]~.
$$

Assume now that all is as in Sect.~\ref{rh}, except that when Clock 1
of a tracer rings, it exchanges energy with the tank
according to the rule in \eref{e:exchange} and not the random-halves rule.
Following the computation in Sect.~\ref{ssa} (details of which are left to the
reader), we see that 
Propositions \ref{p1} and \ref{p2} hold for the present model
provided $\sigma_k$ is replaced by
$$
\sigma_k(\{x_1,\cdot \ldots\cdot, x_k\}, y) \ = \ I_{\{x_1, \ldots, x_k, y \geq 0\}} \
\frac{1}{\sqrt{x_1\cdot \ldots\cdot x_k y}} \ e^{-\beta (x_1+\ldots + x_k +y)}~.
$$
This defines a new family of $\{\hat\mu^{\Tr}\}$ for this model. With 
$\hat\mu^{\Tr}$ in hand, we make the assumption as before
that for $N\gg 1$, the marginals of individual sites have the same form.
Proceeding as in Sect.~\ref{profiles}, we read off the following information on
single cells:
\begin{itemize}
\item[(i)]  stored energy has density ${\rm const.} e^{-\beta y}/\sqrt{y}$ and
mean
${T}/{2}$~;

\item[(ii)] tracer energy has density ${\rm const.}e^{-\beta x}/\sqrt{x}$
and mean ${T}/{2}$~;

\item[(iii)] mean number of tracers, $\kappa = 2\sqrt{\frac{\pi}{T}} \j$~;

\item[(iv)] mean total-cell energy, $E=\frac{T}{2}(1+ \kappa)$~.
\end{itemize}
The rest of the analysis, including (iv), (v), (A) and (B), do not
depend on the local rules (aside from the fact that tracers exiting
a cell have equal chance of going left and right). Thus they
 remain unchanged. Reasoning as in Theorem~\ref{thm1}, we obtain the following:

\begin{proposition}\label{pthm1} Under Assumptions 1 and 2, the profiles for
 the model with energy exchange rule \eref{e:exchange} 
are
\begin{itemize}
\item[$\bullet$] $j(\xi)\ = \ 2(\jl+(\jr-\jl)\xi)~;$
\item[$\bullet$] $Q(\xi) \ = \ 2 (\jl \tl  + (\jr \tr -\jl \tl )\xi)~;$
\item[$\bullet$]$s(\xi )  \ = \ e(\xi) \ = \frac{1}{2}\ Q(\xi)/j(\xi)~;$
\item[$\bullet$] $\kappa(\xi) \ = \ \sqrt{{2\pi}/{s(\xi)}} \ {j(\xi)}/{2}~;$
\item[$\bullet$] $E(\xi)\ = \ s(\xi)+ \kappa(\xi)e(\xi) \ = \ s(\xi) +
  \sqrt{2\pi s(\xi)}\  j(\xi)/2$~.
\end{itemize}
\end{proposition}

Numerical simulations give results in excellent agreement with
these theoretical predictions.


\section{Hamiltonian Models}

In Sect.~\ref{fam}, we introduce a family of Hamiltonian models generalizing
those studied numerically in \cite{Larralde2002,Larralde2003}. A single-cell
analysis similar to that in Section~\ref{stoch} is carried out for this family in
Sect.~\ref{sc2}, and predictions of energy and tracer density profiles are 
made in Sect.~\ref{prof}. We again use the Assumptions in Section~\ref{stoch},
but the predictions here are made under an additional ergodicity assumption,
ergodicity being a property that is  
easy to arrange in stochastic models but not in Hamiltonian ones.
Results of simulations for a specific model are shown in Sect.~\ref{sim2}.
A brief discussion of related models is given in Sect.~\ref{concrete}.


\subsection{Rotating disks models}\label{fam}

We describe in this subsection a family of models quite close to those
studied numerically in \cite{Larralde2002,Larralde2003}. The rules
of interaction (though not the coupling to heat baths) are, in fact,
used earlier in \cite{Rateitschak2000}.

\subsubsection{Dynamics in a closed cell}\label{I}

We treat first the dynamics within individual cells assuming
the cell or box is {\it sealed}, \ie, it is not connected to its neighbors
or to external heat sources.

\medskip
Let $\Gamma_0 \subset {\mathbb R}^2$ be a bounded domain with 
piecewise $C^3$ boundary. In the interior of $\Gamma_0$ lies a
(circular) disk $D$, which we think of as nailed down at its center.
This disk rotates freely, carrying with it a finite amount of kinetic
energy derived from its angular velocity; it will
 play the role of the  ``energy tank" in Sect.~\ref{gsmodm}.
The system below describes the free motion of $k$ point
particles (\ie, tracers) in $\Gamma=\Gamma_0 \setminus D$.
When a tracer runs into $\partial \Gamma_0$, the boundary 
of $\Gamma_0$, the reflection is specular. When it hits
the rotating disk, the energy exchange is according to the rules introduced in
 \cite{Rateitschak2000,Larralde2002,Larralde2003}. 
A more precise description of the system follows.

 The phase space of this dynamical system is
$$
\bar \Omega_k = (\Gamma^k \times \partial D \times {\mathbb
  R}^{2k+1})/\sim~
$$
where

${\bf x} = (x_1, \dots, x_k) \in \Gamma^k$ denotes the positions of the $k$ tracers,

$\theta \in \partial D$ denotes the angular position of a (marked) point on
the boundary of the turning disk,

${\bf v} = (v_1, \dots, v_k) \in {\mathbb R}^{2k}$ denotes the 
velocities of the $k$
tracers,

$\omega \in \mathbb R$ denotes the angular velocity of the turning disk,

\medskip
\noindent and $\sim$ is a relation that identifies pairs of points in
the collision manifold $M_k = \{({\bf x}, \theta, {\bf v}, \omega)
 : x_\ell \in \partial \Gamma$ for some $\ell\}$.
The rule of identification is given below.

The flow on $\bar \Omega_k$ is denoted by $\bar \Phi_s$. 
As long as no collisions are involved, we have
$$
\bar \Phi_s\ ({\bf x}, \theta, {\bf v}, \omega) \ = \ 
({\bf x}+s{\bf v},  \theta+ s \omega, {\bf v},\omega)~.
$$
We assume at most one tracer collides with $\partial \Gamma$
at any one point in time. ($\bar \Phi_s$ is not defined at multiple collisions,
which
occur on a set of measure zero.) At the point of impact,
\ie, when $x_\l  \in \partial \Gamma$ for one of the $\l$,
let $v_\l =(v_\l ^{\rm t}, v_\l ^{\rm n})$ be the tangential and normal
components of $v_\l $.
What happens subsequent to impact depends on whether $x_\l  \in
\partial \Gamma_0$ or $\partial D$.  In the case of a collision with
$\partial \Gamma_0$,
the tracer bounces off $\partial \Gamma_0$ with angle of reflection equal
to angle of incidence, \ie, 
\begin{equation}\label{mml0}
(v_\l ^{\rm n})' = -v_\l ^{\rm n}, \qquad (v_\l^{\rm t})' = v_\l ^{\rm t} ~,
\end{equation}
and the other variables are unchanged.
In the case of a collision with the disk,
the following {\em energy exchange} takes place 
between the disk and the tracer:
\begin{equation}\label{mml}
(v_\l ^{\rm n})'  =  - v_\l ^{\rm n}, \qquad
(v_\l ^{\rm t})'  =  \omega\ , \qquad
\omega'  =  v_\l ^{\rm t}\ .
\end{equation}
Here we have, for simplicity, taken the radius of the disk, the moment
of inertia of the disk,
and the mass of tracer in such a way that the coefficients in
Eq.~\eref{mml} are equal to 1. 
The identification in the definition of $\bar \Omega_k$ is 
$z \sim z'$ where $z, z' \in M_k$ are such that all of their
coordinates are equal except that $v_\ell$ and $\omega$ in $z$
are replaced by the corresponding quantities with primes in
Eqs.~\eref{mml0} and \eref{mml} for $z'$. We also write $F(z)=z'$.

\medskip
Note that in both \eref{mml0} and \eref{mml}, total energy is conserved, \ie, $|v|^2+\omega^2=|v'|^2+(\omega')^2$.
The energy surfaces in this model are therefore
$$
\bar \Omega_{k,E} \ = \ (\Gamma^k \times \partial D \times
S^{2k}_E)/\sim
$$
where
$$S^{2k}_E=\{(v_1, \dots, v_k, \omega) \in {\mathbb R}^{2k+1}:
\sum |v_\l |^2 + \omega^2=E\}~.
$$

\smallskip
We claim that the natural invariant measure, or Liouville measure,
of the (discontinuous) flow $\bar \Phi_s$ on $\bar \Omega_k$ is 
$$
\bar m_{k} = (\lambda_2|_\Gamma)^k \times (\nu_1|_{\partial D})
 \times \lambda_{2k+1}
$$
where $\lambda_d$ is
$d$-dimensional Lebesgue measure and $\nu_d$ is surface area on the
relevant $d$-sphere. Once the invariance of $\bar m_k$
 is checked, it will follow immediately that the induced 
measures $\bar m_{k,E} = (\lambda_2|_\Gamma)^k \times (\nu_1|_{\partial D}) 
\times \nu_{2k}$ on $\bar \Omega_{k,E}$ are $\bar \Phi_s$-invariant,
as are all measures on $\bar \Omega_k$ of
the form $\psi(E) \bar m_{k,E}$ for some $\psi:[0,\infty) \to [0,\infty)$.

The invariance of $\bar m_k$  is obvious away from collisions
and at collisions with $\partial \Gamma_0$. Because collisions occur 
one at a time, it suffices to consider a single collision 
between a single tracer and the disk. The problem, therefore, is reduced
to the following: Consider $\bar \Phi_s$ on $\bar \Omega_1$, and
let $M_{1,D}$ denote the part of the collision manifold involving $D$.
To prove that $\bar m_1$ is preserved in a collision with $D$,
if suffices to check that for $A \subset M_{1,D}$ and $\varepsilon>0$
arbitrarily small,
$$\bar m_1\big(\cup_{-\varepsilon<s<0} \Phi_s(A)\big)
= \bar m_1\big(\cup_{0<s<\varepsilon} \Phi_s(F(A))\big)~.
$$
 We leave this as a calculus exercise.


\subsubsection{Coupling to neighbors and heat baths}\label{II}

We now consider a chain of $N$ identical
copies of the dynamical system described
in Sect.~\ref{I}, and define couplings between nearest neighbors 
and between end cells and heat baths.

\medskip

\begin{figure}[ht]
\begin{center}
\epsfig{file=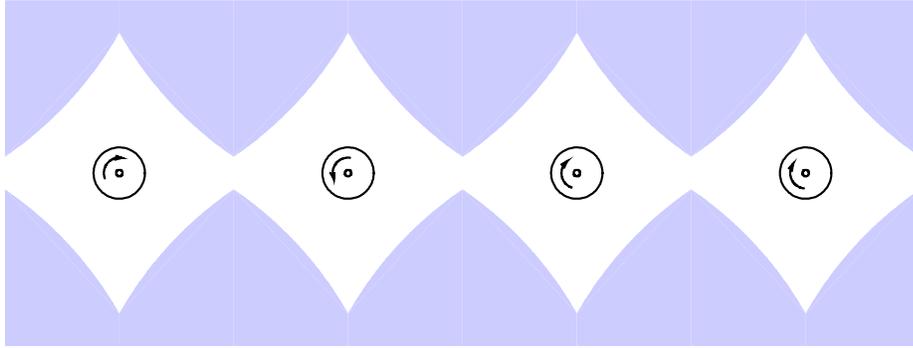,width=0.8\hsize}
\end{center}
\caption{A row of diamond-shaped boxes with small lateral holes 
(made by removing vertical walls corresponding to 
$\gammar^{(i)}=\gammal^{(i+1)}$) to 
allow the tracers to go from one box to the next. 
The shapes of the ``boxes" can be quite general for much of our theory.
The configuration shown 
is the one used in the simulations discussed in
Sect.~\ref{sim2}, but with larger holes for better visibility.}\label{f:cells}
\end{figure}

Let $\gammal $ and $\gammar $ be two marked subsegments of
$\partial \Gamma_0$ of equal length; these segments will serve as 
openings to allow tracers to pass between cells. For now it is best to think of
$\Gamma_0$ as having a left-right symmetry, and to think of $\gammal$
and $\gammar$ as vertical and symmetrically placed (as in Fig.~\ref{f:cells}), 
although as we will see, these geometric details are not relevant
for the derivation of mean energy and tracer profiles.
We call the segments $\gammal $ and $\gammar $ in the $i$th cell
$\gammal^{(i)}$ and $\gammar^{(i)}$.
For $i=1, \dots, N-1$, we identify $\gammar^{(i)}$ with $\gammal^{(i+1)}$,
that is to say, we think of the domains of the $i$th cell and the $(i+1)$st
cell
as having a wall in common, namely $\gammar^{(i)}$ and $\gammal^{(i+1)}$,
and {\it remove} this wall, so that tracers that would have collided with
it simply continue in a straight line into the adjacent cell. 
(See Fig.~\ref{f:cells}.)

\medskip
Tracers are injected into the system as follows.  Consider, for example,
the bath on the left. We say the injection rate is $\rate$ if at the ring of
an exponential clock of rate $\rate$, a single tracer enters  cell $1$
via $\gammal^{(1)}$. (Note that the rate $\rate$ is not the injection rate 
per unit length of the opening $\gammal^{(1)}$ but {\it per unit time}.)
The points of entry and velocities of entering tracers are {\it iid},
the law being the one governing the collisions of tracers
with $\gammal^{(1)}$.  That is to say, the point of
entry is uniformly distributed on $\gammal^{(1)}$, and the velocity $v$
has density
\begin{equation}\label{e:density}
c \  e^{-\beta |v|^2} \ |v| |\sin(\varphi)| \ dv\ , \qquad c =
\frac{2 \beta^{3/2}}{\sqrt \pi}~,
\end{equation}
where $v \in {\mathbb R}^2$ points into $\gammal^{(1)}$ and 
$\varphi \in (0, \pi)$ is the
angle $v$ makes with $\gammal^{(1)}$ at the point of entry. 
(This is the distribution of $v$ at collisions for particles
 with velocity distribution   $\frac{\beta}{\pi} \exp(-\beta |v|^2)
 dv$.)
Here $\beta =1/T$ where $T$ is said to be the temperature 
of the bath. Observe that the mean energy
of the tracers injected into the system by a bath at temperature $T$ is
not $T$ but $3T/2$. Injection from the right is done similarly,
via the opening $\gammar^{(N)}$.
When a tracer in the chain reaches $\gammal^{(1)}$ or 
$\gammar^{(N)}$, it  vanishes into the baths.

\medskip
This completes the description of our models. We remark that
the process above is
a Markov process in which
the only randomness comes from the action of the baths.
Once a tracer
is in the system, its motion is governed by rules that are
entirely deterministic.


\subsection{Single-cell analysis}\label{sc2}

In analogy with Sect.~\ref{ssa}, we investigate in this
subsection the invariant measure
for a single cell  coupled to two  heat baths with parameters $T$ and $\rate$.

Let  $\bar \Omega_k$ and $\bar \Omega_{k,E}$ be as in Sect.~\ref{I}.
As before, a state of this system is represented by a point in
$$
\Omega \ = \ \cup_{k=0}^\infty \Omega_k \ = \ \cup_{k=0}^\infty
\cup_{E \geq 0} \Omega_{k,E} 
$$
where $\Omega_k$ and $\Omega_{k,E}$ are quotients of $\bar \Omega_k$
and $\bar \Omega_{k,E}$ respectively obtained by identifying 
permutations of the $k$ tracers.
With $\{ \cdots\}$ representing unordered sets
as before, points in $\Omega$ are denoted by
$z=(\{x_1, \dots, x_k\}, \theta; \{v_1, \dots, v_k\}, \omega)$ or simply
$(\{x_\l \}, \theta; \{v_\l \}, \omega)$, with $v_\l $ understood
to be attached to $x_\l $. The quotient measures of $\bar m_k$ and
$\bar m_{k,E}$ are respectively $m_k$ and $m_{k,E}$.

Abusing notation slightly, we continue to use $\bar \Phi_s$ to denote 
the semi-flow on $\bar \Omega$,
and let $\Phi_s$ denote the induced semi-flow on $\Omega$.
Then $\Phi_s$ is as in Sect.~\ref{I} except where tracers exit or enter
the system. More precisely,
if  $\Phi_s(z) \in \Omega_k$ for all $0 \leq s < s_0$, and
a tracer exits the system at time $s_0$, then 
$\Phi_{s_0}(z)$ jumps to $\Omega_{k-1}$. 
Similarly, if a tracer is injected from one of
the baths at time $s_0$, then instantaneously $\Phi_{s_0}(z)$ jumps to
$\Omega_{k+1}$, the destination being given by a probability
distribution.

Let $|\gamma|$ denote the length of the segment $\gammal $ or $\gammar
$.

\begin{proposition}\label{mu} There is an invariant probability
  measure $\mu$ with the following properties:

\medskip
\noindent (a) the number of tracers present is a Poisson random variable
with mean $\kappa $ where
$$
\kappa = {2\sqrt{\pi}}\,\frac{\lambda_2(\Gamma)}{|\gamma|} \frac{\rate}{\sqrt
T}~;
$$
(b) the conditional density of $\mu$ on $\Omega_k$ is $c_k 
\sigma_k dm_k$ where
$$
\sigma_k(\{x_\l \}, \theta; \{v_\l \}, \omega) \ = \ e^{-\beta( \omega^2 +
\sum_{\l=1}^k |v_\l |^2)}
$$

and $c_k$ is the normalizing constant.
\end{proposition}

We observe as before that the Poisson parameter $\kappa$ is 
proportional
to $\rate$ (the higher the injection rate, the more tracers in the cell) and
inversely proportional to $\sqrt T$, \ie,  the speed of the tracers
(the faster the tracers, the sooner they
leave). Unlike the models considered in Sect.~\ref{stoch}, where the tracers
are assumed
to leave the cell at a rate equal to their speed, here the ratio
$\lambda_2(\Gamma)/|\gamma|$
appears, as it should: the smaller the passage way, the longer it takes for the
tracers to leave.

Notice that we have not claimed that $\mu$ is unique.

\medskip
We introduce some notation in preparation for the proof.
For $A \subset \Omega_k$ and $h>0$, we let
$\Phi_{-h}(A)$ denote the set of all initial states in $\Omega$ that in time
$h$ evolve into $A$ {\it assuming no new tracers are injected into the system
between times $0$ and  $h$}.\footnote{Notice that (1)
$\{\Phi_h, h \geq 0\}$ is a semi-flow, and $\Phi_{-h}$ is not
defined; (2) $\Phi_{-h}(A)$ as defined is $\neq (\Phi_h)^{-1}(A)$.}
Then $\Phi_{-h}(A) =\cup_{n \geq 0} \Phi_{-h}^{(n)}(A)$ where
$\Phi_{-h}^{(n)}(A) = \Phi_{-h}(A) \cap \Omega_{k+n}$, \ie,
$\Phi_{-h}^{(n)}(A)$ is the set of states where initially $k+n$ tracers
are present, and by the end of time $h$ exactly $n$ of these tracers
have exited and the remaining $k$ are described by a state in $A$.

\begin{lemma}\label{oldl31} Let $\mu$ be as in Proposition~\ref{mu}, 
and let
$A_\varepsilon$ be a cube of sides $\varepsilon$ in $\Omega_k$\,,
$\varepsilon$ small enough that
$\mu(A_\varepsilon) \approx p_k c_k \sigma_k(\bar z) \varepsilon^{4k+2}$
for some $\bar z$.
We assume the following holds for all small  $h>0$:

(i) no tracers are injected into the system on the time interval $(0,h]~;$

(ii) $\Phi_h(\Phi_{-h}^{(0)}(A_\epsilon )) = A_\epsilon$.

\noindent Then
\begin{equation}\label{e42} \mu(\Phi_{-h}(A_\epsilon ))\,=\, \sigma_k(\bar z)
\varepsilon^{4k+2} 
\left(p_k c_k + 2h \frac{|\gamma|}{c} p_{k+1} c_{k+1} \ + \ \oo(h) \right)
\end{equation}
where $p_k=\mu(\Omega_k)$ and $c = \frac{2}{\sqrt \pi} \beta^{{3}/{2}}$.
\end{lemma}

\noindent {\bf Proof:} The idea is that for a particle to exit in the
very short time
$h$, it must be close to the exit $\gammal$ or $\gammar$ and move
towards it without colliding with the disk or the boundary, {\em or}
it must have very large speed (and that is improbable).

\medskip
By assumption (i), we have
$\mu(\Phi_{-h}(A_\epsilon )) = \sum_{n \geq 0} 
\mu( \Phi_{-h}^{(n)}(A_\epsilon))$. The  $n=0$ term is handled
easily: By virtue of (i) and (ii), the situation is equivalent 
to that in Sect.~\ref{I}. Since $\mu|_{\Omega_k}$ is invariant for
the closed dynamical system with $k$ tracers, we
have $\mu(\Phi_{-h}^{(0)}(A_\epsilon ))=\mu(A_\epsilon )$. 

\medskip
Consider next $n=1$. We give the estimate for 
$\mu( \Phi_{-h}^{(1)}(A_\epsilon))$ assuming $\gammal$ 
and $\gammar$ are straight-line segments, leaving the general case
(where these segments may be curved) to the reader. 
First some notation:
For $v \in {\mathbb R}^2$
and $a>0$, let $E(v, a, \L)$ be the parallelogram on the same
side of  $\gammal $ as $\Gamma$ and with the property that 
one of its sides is $\gammal $ while the other is parallel to $v$ 
and has length $a$; $E(v, a, \R)$ is defined similarly. 
To simplify the discussion, we assume that for $a>0$ sufficiently
small, $E(v, a, \L)$ 
and $E(v, a, \R)$ are contained in $\Gamma$, and leave to the reader
the verification that ``corners'' at the end of $\gamma_\L$ or
$\gamma_\R$ lead to higher order terms (in the variable $h$ used below).

Starting from a state in $\Phi_{-h}^{(1)}(A_\epsilon)$, 
we let $x$ and $v$ denote the initial position and velocity of the tracer that 
exits before time $h$, and treat separately the cases
(1) $|v| \leq \frac{a}{h}$ and (2) $|v| > \frac{a}{h}$.
In Case (1), in order for the tracer to exit before time $h$, we must
have $x \in E(v,h|v|,\L)\cup E(v,h|v|,\R)$, and $v$ must point toward 
the exits. Since $\lambda_2(E(v,h|v|,\L))=\lambda_2(E(v,h|v|,\R))= 
h|v| |\sin(\varphi)||\gamma|$ where $\varphi$ is the angle $v$ 
makes with $\gammal$ or $\gammar$, we obtain
\begin{eqnarray*}
\mu( \Phi_{-h}^{(1)}(A_\epsilon) \cap \{|v| \leq \frac{a}{h}\}) & = &
\sigma_k(\bar z) \varepsilon^{4k+2} p_{k+1} c_{k+1} 
\cdot 2h |\gamma| \int_0^\pi d\varphi
|\sin(\varphi)| \ \int_{|v| \leq \frac{a}{h}}  dv |v| e^{-\beta |v|^2}\\
& = & \sigma_k(\bar z) \varepsilon^{4k+2} p_{k+1} c_{k+1} 
\cdot 2h |\gamma| \frac{1}{c} (1 + \oo(h))
\end{eqnarray*}
with $c=2\beta ^{3/2}/\sqrt{\pi}$.
For Case (2), we have the trivial estimate 
$\sigma_k(\bar z) \varepsilon^{4k+2} p_{k+1} c_{k+1} 
\cdot \oo(h)$.

\medskip
To see that the terms corresponding to $n>1$ are negligible, we
first derive the bound
\begin{equation}\label{e:44}
\mu(\Phi_{-h}^{(n)}(A)) \ \leq \ \sigma_k(\bar z) \varepsilon^{4k+2}
p_{k+n} c_{k+n} \ ( 2h |\gamma| \frac{1}{c}  + \oo(h) )^n~.
\end{equation}
Then we compute the growth rate of $p_{k+n} c_{k+n}$.
By the definitions of these numbers, we have
$$
\frac{c_{k+1}}{c_k} \cdot \frac{p_{k+1}}{p_k} \ = \
\left(\frac{\lambda_2(\Gamma)}{k+1} \int_{{\mathbb R}^2} 
e^{-\beta |v|^2} dv \right)^{-1}
\left( \frac{1}{k+1} \ 2 \sqrt \pi \frac{\lambda_2(\Gamma)}{|\gamma|}
\frac{\rate }{\sqrt{ T}} \right)~,
$$
giving
\begin{equation}\label{e:45}
p_{k+n} c_{k+n} \ = \ \left(\frac{c \rate}{|\gamma|} \right)^n \ 
p_k c_k, \qquad n \geq 1 ~.
\end{equation}
{}From \eref{e:44} and \eref{e:45} it follows that 
$\mu(\Phi_{-h}^{(n)}(A)) \leq \sigma_k(\bar z) \varepsilon^{4k+2} 
({\rm const.} \cdot h)^n$. 

\medskip
The asserted bound \eref{e42} for $\mu(\Phi_{-h}(A))$ is proved.
 \hfill $\square$

\vskip .2in

The main difference between the proofs of Propositions~\ref{p1} and \ref{mu}
is that Hamiltonian models have both geometry and memory.
In preparation for the proof, we introduce the following language.
Let $A_\varepsilon$ be as in Lemma~\ref{oldl31}. For $\ell=1,\dots, k$,
we let $X_\ell$ denote the projection of $A_\varepsilon$ onto the plane
of its $x_\ell$-coordinate, and $V_\ell$ the projection of
$A_\varepsilon$ onto the plane of its $v_\ell$-coordinate 
(so that $X_\ell$ and $V_\ell$ are $\varepsilon$-squares in
$\Gamma$ and ${\mathbb R}^2$ respectively).
We assume for simplicity that for each $\ell$, either $X_\ell$ is 
a strictly positive distance from $\gammal $ and $\gammar $, 
in which case we say $X_\ell$ is {\it in the interior},
or one of its sides is contained in $\gammal $ or $\gammar $. 
In the latter case, we say $X_\l $ is {\it adjacent to an exit}. 
We further assume that if $X_\l $ is adjacent to an exit, then 
 either all $v_\l  \in V_\l$ point toward the exit or away from it.

\bigskip
\noindent {\bf Proof of Proposition~\ref{mu}:} \   The invariance of 
$\mu$ is already noted in Sect.~\ref{I} except where it pertains to 
entrances and exits of tracers. We focus therefore on these events,
noting that the probability of more than one tracer entering
on the time interval $(0,h)$ is $\oo(h)$, as is the probability of a
tracer entering and leaving (immediately) on this time interval.
These scenarios will be ignored.

Let $A_\varepsilon$ be as above. We seek to show as before that
$\frac{d}{dh} \int  I_{A_\varepsilon}(z') P^h(dz'|z) \mu(dz)|_{h=0} =0$.
Here it is necessary to treat separately the following configurations 
for $A_\varepsilon$:

\medskip
\noindent {\bf Case 1.} The following holds for all $\ell$: $X_\ell$ can
be in the interior or adjacent to an exit, and if it is adjacent to an exit,
then all $v_\ell$ in $V_\ell$ must point toward the exit. 
Notice that
this configuration is relatively inaccessible, meaning the probability
of a new tracer entering on $(0,h)$ leading to a state in
$A_\varepsilon$ is $\oo(h) \mu(A_\varepsilon)$. Notice also that
this configuration has the property 
$\Phi_h(\Phi_{-h}^{(0)}(A_\epsilon )) = A_\epsilon$, so that 
the contribution of the no-new-tracers event to
$\int I_{A_\varepsilon}(z') P^h(dz'|z) \mu(dz)$ is,
by Lemma~\ref{oldl31},
\begin{equa}[3][e:46]
& \ & (1-h\rate)^2 &\ \sigma_k(\bar z) \varepsilon^{4k+2}\ \bigl(p_k c_k + 2h
|\gamma| \frac{1}{c} p_{k+1} c_{k+1} + \oo(h)\bigr)\\
& = &\ \  \sigma_k(\bar z) \varepsilon^{4k+2} &\ \left(p_k c_k (1-2h\rate) +
2h |\gamma| \frac{1}{c} p_{k+1} c_{k+1} + \oo(h) \right)\\
& = & \sigma_k(\bar z) \varepsilon^{4k+2}& \ \left(p_k c_k + \oo(h) \right)~,\\
\end{equa}
the last equality being valid on account of Eq.~\eref{e:45}.

\medskip
\noindent {\bf Case 2.} $X_1$ is adjacent to an exit and $v_1$
points away from it; $X_\ell$ and $V_\ell$  for
$\ell >1$ are as in Case 1. In this configuration, there is a part of
$X_1$ that can only 
be reached in time $h$ if one starts from outside. This
region is a parallelogram similar to that in the proof of
Lemma~\ref{oldl31} but with one of its sides equal to 
$X_1 \cap \gammal $ or $X_1 \cap \gammar $.
Following the estimates in Case 1, we obtain that 
the contribution of the no-new-tracers event to
$\int I_{A_\varepsilon}(z') P^h(dz'|z) \mu(dz)$ in this case is
\begin{equation}\label{e:488}
\sigma_k(\bar z) \varepsilon^{4k+2} \ p_k c_k \left(1  -
\frac{h}{\varepsilon} |\bar v_1| |\sin(\bar \varphi_1)| +\oo(h)\right)
\end{equation}
where $\bar v_1$ is the $v_1$ coordinate of $\bar z$ and
$\bar \varphi_1$ is the angle $\bar v_1$ makes with
$\gammal $ (or $\gammar $).

We now argue that the negative term above is balanced by the
contribution of the event in which a new tracer enters on the time
interval $(0,h)$. This new tracer must have
$v_1 \in V_1$ and must enter through the $\varepsilon$-segment
$X_1 \cap \gammal $ or $X_1 \cap \gammar $. We claim that the
probability of this event is
\begin{equation}\label{e:47}
p_{k-1} c_{k-1} \sigma_k(\bar z) e^{\beta |\bar v_1|^2} \varepsilon^{4k-2}  
\ \cdot \ \rate h \frac{\varepsilon}{|\gamma|} \ \cdot \ 
c |\sin(\bar \varphi_1)| |\bar v_1| e^{-\beta |\bar v_1|^2} \varepsilon^2~.
\end{equation}
The first factor in \eref{e:47} is the $\mu$-measure of the states 
corresponding to those in $A_\varepsilon$ but {\it without} the tracer 
with position and velocity $(x_1, v_1)$; the second factor is 
the probability of a tracer entering through the designated segment, 
and the third is the fraction of tracers entering with velocity $\in V_1$ 
(see \eref{e:density}). That \eref{e:488} and \eref{e:47} add up to
$\mu(A_\varepsilon)(1+\oo(h))$ again follows from \eref{e:45}.

\medskip
\noindent {\bf Case 3.} $X_1$ and $X_2$ are adjacent to exits,
$v_1$ and $v_2$ point away from the exits in question,
and  $X_\ell$ and $V_\ell$ are 
as in Case 1 for $\ell >2$. We assume for simplicity that
either $(X_1 \times V_1) \cap (X_2 \times V_2)= \emptyset$ or
$X_1 \times V_1 = X_2 \times V_2$. 

In the case $(X_1 \times V_1) \cap (X_2 \times V_2)= \emptyset$,
the contribution of the no-new-tracers event  is
\begin{equation}
\sigma_k(\bar z) \varepsilon^{4k+2} \ p_k c_k \left(1  -
\frac{h}{\varepsilon} |\bar v_1| |\sin(\bar \varphi_1)| 
- \frac{h}{\varepsilon} |\bar v_2| |\sin(\bar \varphi_2)| +\oo(h)\right)~,
\end{equation}
and this is cancelled perfectly by the estimate corresponding to \eref{e:47}.

In the case $X_1 \times V_1 = X_2 \times V_2$,
on $\bar \Omega_k$, where tracer positions and velocities are
regarded as ordered $k$-tuples, the set of states where
both $(x_1, v_1)$ and $(x_2, v_2)$ are not reachable in time $h$ is $\oo(h)$,
and the set where exactly one of these is not reachable is the union of
two sets that project to the same set under $\pi_k$. Thus the estimates
for both cases are as in Case 2.

\medskip
The remaining cases are handled similarly.   \hfill $\square$

\smallskip
\begin{proposition} For the $N$-chain defined in Sect.~\ref{II}
with $\tl =\tr =T$ and $\jl =\jr =\j$,
the $N$-fold product $\mu \times \dots \times \mu$ is invariant.
\end{proposition}

It suffices to check that the transfer of energy from one cell to the next
leads to the correct relation between $p_k c_k$ and $p_{k+1} c_{k+1}$.
The proof is left to the reader.


\subsection{Derivation of equations of macroscopic profiles}\label{prof}

Having found the candidate family of Gibbs measures $\{\mu^{\Tr}\}$,
we now proceed as in Sect.~\ref{profiles}, seeking to derive  the relevant 
macroscopic profiles under Assumptions 1 and 2; see Sect.~\ref{profiles}. 
There are two new problems, leading to two additional assumptions
which we now discuss.

\medskip
The first problem is that of uniqueness and ergodicity. 
Unlike their stochastic counterparts, the 
Hamiltonian chains defined in Sect.~\ref{fam} may not be ergodic;
they are, in fact, easily shown to be nonergodic for certain choices 
of $\Gamma_0$. Without ergodicity, it is not clear 
how to make sense of the notion  of {\it local temperature}, which 
lies at the heart of Assumption 2. Postponing a discussion till later, 
we bypass this issue by introducing

\medskip
\begin{changemargin}{1cm}{1cm} 
{\bf Assumption 1'.}  {\it We assume $\mu_N$ is the unique invariant
probability measure for the $N$-chain 
defined in Sect.~\ref{fam}. It follows that $\mu_N$ is ergodic.}
\end{changemargin} 

\medskip
Our next assumption pertains to the asymmetry of exit distributions
from each cell in the finite chain. Let $j_{N,i}$ denote the mean number
of exits out of the $i$th cell per unit time in the $N$-chain. (Each time 
a tracer exits the $i$th cell, it is counted as ``an exit", even if it is the 
same tracer that re-enters and exits again.) Then 
$j_{N,i}=j_{N,i,\L} + j_{N,i,\R}$ where $j_{N,i,\L}$ and $j_{N,i,\R}$ are 
the numbers of exits per unit time that go to the $(i-1)$st and $(i+1)$st
cells respectively. In the random-halves model studied in Section \ref{stoch}, 
our local rule is that $j_{N,i,\L}=j_{N,i,\R}$. That is typically {\it not} the case
in models in which local rules are purely dynamical, such as the 
Hamiltonian models under consideration. Assumption 3 controls this deviation
from symmetry for $j_{N,i}$ and $Q_{N, i}$ where $j_{N,i}$ is as above
and $Q_{N, i}$ denotes
the mean total energy transported out of the $i$th cell per unit time 
in the $N$-chain.

\medskip
\begin{changemargin}{1cm}{.9cm} 
{\bf Assumption 3.}  {\it For each cell configuration
$(\Gamma_0,\gammal, \gammar)$ and parameters $\tl$, $\tr$, $\jl$, $\jr>0$,
we assume
\begin{itemize}
\item[(i)]
there exists an $\alpha \ge 0$ such that for all large $N$, the following
hold for all $i$:
\begin{equa}
|j_{N,i,R}- \frac{1}{2}j_{N,i}|  \leq & \frac{\alpha}{N}~;   
\quad |Q_{N,i,R}- \frac{1}{2}Q_{N,i}| \leq \frac{\alpha}{N}~;\\
|(j_{N,i,\R}- \frac{1}{2}j_{N,i}) & - (j_{N,i+1,\R}-  \frac{1}{2} j_{N,i+1})| 
\leq  \frac{\alpha}{N^2}~;\\
|(Q_{N,i,\R}-  \frac{1}{2}Q_{N,i}) & - (Q_{N,i+1,\R}-  \frac{1}{2}Q_{N,i+1})| 
\leq \frac{\alpha}{N^2}~;
\end{equa}
\item[(ii)] as $N \to \infty$, the profiles $j_{N,i}$ and $Q_{N,i}$ tend to
$\CC^2$ functions $j$ and $Q$ on $[0,1]$.
\end{itemize}
}
\end{changemargin} 

\medskip
\begin{remark}\label{a3}\rm Assumption 3 is consistent with the following
observations: For a cell in the $N$-chain, the temperature difference between
the cell on its left and the one on its right is of order $|\tl-\tr|/N$, so one
expects the marginal of $\mu_N$ at this site to deviate from the equilibrium
measure in Sect~\ref{sc2} by the same order of magnitude. This deviation
is in turn reflected in the differences $|j_{N,i,\R}-j_{N,i,\L}|$ and
$|Q_{N,i,\R}-Q_{N,i,\L}|$. Similarly, if the second differences
are well behaved as we assume, 
their orders of magnitude as indicated above are dimensionally correct. 
Detailed dependencies of this asymmetry 
on the physical parameters are beyond the scope of this 
paper.\footnote{We thank H. Spohn for
  interesting correspondence on this point.}
\end{remark}

We are primarily interested in
situations where $\alpha\ll 1$, which happens when there is good mixing
{\it within individual cells}. Good geometry of $\Gamma_0$ (such as concave
walls and the absence of ``traps") and small passageways between adjacent cells 
(so most tracers stay in the cell for a long time) are conducive to fast mixing 
within individual cells. The presence of large numbers of tracers also enhances 
mixing. \footnote{It is important to distinguish between the following two levels
of mixing: mixing {\it within cells}, and mixing {\it in the chain}.
For example, small passageways between cells enhance mixing 
of the first kind but are obstructions to the latter.}

\begin{theorem}\label{thm2} Under Assumptions 1, 1', 2 and 3, 
we have the following limiting profiles for the models in Sect.~\ref{fam}
as $\alpha$ in Assumption 3 tends to $0$.
\SI{\ $\bullet$}
\begin{itemize}
\item[$\bullet$] mean number of exits out of a site
per unit time~: 
$$
j(\xi) \ = \ 2\left( \jl   + (\jr -\jl )\xi\right )~;
$$
\item[$\bullet$] mean total energy transported out of a site
per unit time~:
$$
Q(\xi) \ = \ 3\bigl(\jl \tl  + (\jr \tr -\jl \tl )\xi\bigr)~;
$$
\item[$\bullet$] mean stored energy at a site~:
$$
s(\xi) \ = \ \frac{1}{3} \ \frac{Q(\xi)}{j(\xi)} \ = \
\frac{1}{2} \frac{\jl \tl  + (\jr \tr -\jl \tl )\xi}{\jl +(\jr -\jl )\xi}~;
$$
in the case $\jl =\jr $, this simplifies to
$s(\xi) =  \frac{1}{2} \bigl(\tl  + (\tr -\tl )\xi\bigr)$~.
\item[$\bullet$] mean tracer energy~: $e(\xi)=2 s(\xi)~;$
\item[$\bullet$] mean number of tracers~:
$$
\kappa(\xi) \ = \ \frac{\lambda_2(\Gamma)}{|\gamma|} \ \sqrt{\frac{\pi}{2s(\xi)}} \
j(\xi) \
$$
where $|\gamma|=|\gammal| = |\gammar|$ is the size of the passage
between adjacent cells\,$;$
\item[$\bullet$] mean total-cell energy~:
$$
E(\xi)\ = \ s(\xi) + \kappa(\xi)e(\xi) \ = \ s(\xi)  +
\frac{\lambda_2(\Gamma)}{|\gamma|}
\sqrt{2\pi s(\xi)}\  j(\xi) ~.
$$
\end{itemize}
\end{theorem}

\noindent {\bf Proof:} The proof follows that of Theorem \ref{thm1}.
First we read off the pertinent information from Proposition~\ref{mu} for
a single cell connected to two heat baths with parameters $T$ and $\rate$:
\begin{itemize}
\item[(i)]  stored energy has density $\frac{\sqrt{\beta }}
{\sqrt{\pi y}} e^{-\beta y}$
and mean $s=\frac{T}{2}~;$

\item[(ii)] tracer energy has density $ \beta e^{-\beta x}$ and mean 
$T~;$\footnote{Note that this is the energy density when the tracers
are {\it in the box}, to be distinguished from (vi).}

\item[(iii)] mean number of tracers, $\kappa = 
\frac{\lambda_2(\Gamma)}{|\gamma|} \sqrt \pi \frac{2 \rate }{\sqrt T}~;$

\item[(iv)] mean total-cell energy, $E=T(\kappa + \frac{1}{2})~;$

\item[(v)] mean number of jumps out of cell per unit time, $ j =2 \rate~;$

\item[(vi)] mean total energy transported out of cell per unit time, $Q= 
\frac{3T}{2} \cdot j = 3T \j$~.
\end{itemize}
To prove (i), for example, we condition on the event that exactly $k$ tracers
are present. Integrating out all other variables, we obtain that the
distribution of $\omega$ is ${\rm const.} e^{-\beta \omega^2}$. Thus the
distribution of $s=\omega^2$ is as claimed. Items (ii) -- (iv) are proved similarly,
and (v) and (vi) are deduced from the fact that the cell is in equilibrium
with the two baths.

\medskip
As in the proof of Theorem \ref{thm1}, we deduce next the profiles of
$j$ and $Q$. Here,
$$
j_{N,i,\R} - j_{N, i+1, \L} \ = \ j_{N, i-1, \R} - j_{N,i,\L}~.
$$
This together with Assumption 3 implies that as $N \to \infty$, 
the limiting profile $j(\xi)$ has the property $|j''|={\cal O}(\alpha)$. 
Since $j(0)=2\jl$ and $j(1)=2\jr$, we have
$j(\xi) \to 2( \jl   + (\jr -\jl )\xi)$ as $\alpha \to 0$. Analogous reasoning 
gives $Q(\xi) \to 3(\jl \tl  + (\jr \tr -\jl \tl )\xi)$ as $\alpha \to 0$.

\medskip
To deduce the remaining profiles, consider $\xi \in (0,1)$. We let 
$\mu_{N, [\xi N]} \to \mu^{T(\xi), \rate(\xi)}$, and use the single-cell 
information above combined with the profiles of $j$ and $Q$ to identify 
$T(\xi)$ and $\rate(\xi)$. For example, the formula for $s$ is 
obtained as follows: $T=\frac{2}{3}Q/j$ is from (v)
and (vi), and $s =\frac{1}{2}T$ is from (i).
\hfill  $\square$


\subsubsection{Comparisons of models}

\noindent 1. {\it Predicted profiles for Hamiltonian and stochastic models}. \
We observe that as functions,  the predicted formulas in Theorem~\ref{thm2} are
of the same type as their counterparts in Theorem~\ref{thm1} but the constants
are different. The similarity stems from the fact that they are derived from
the same general principles. The differences in constants reflect the
differences in $\mu^{\Tr}$, which in turn reflect the differences in local
rules
(see below).

\medskip
\noindent 2. {\it Relation between $s$ and $e$}. \  To highlight the role
of the local rules in the profiles studied in this paper, we recall the
relation
between stored energy $s$ and individual tracer energy in the various models
encountered:
\begin{itemize}
\item[(a)] Random halves (Sects.~\ref{rh}--\ref{profiles}): $s=2e$. (At
collisions, energy
is split evenly on average, but the expected time for the next clock is
longer for slower tracers.)
\item[(b)] Stochastic models simulating Hamiltonian systems where both disk and
tracer have a single degree of freedom (Sect.~\ref{otherexamples}): $s=e$.
\item[(c)] Hamiltonian models in which the disk has one degree of freedom
and tracers have two (Section~\ref{fam}): $e=2s$.
\end{itemize}
To this list, we now add one more example, namely
\begin{itemize}
\item[(d)] Hamiltonian models in which the disk has one degree of freedom
and tracers
have $3$: Consider the model described in Sect.~\ref{fam}, but with
$\Gamma_0 \subset {\mathbb R}^3$ and the disk replaced by a cylinder
that rotates along a fixed axis. Here, Liouville measure for a closed system
with $k$ tracers is 
$\bar m_{k} = (\lambda_3|_\Gamma)^k \times (\nu_1|_{\partial D})
 \times \lambda_{3k+1}$
({\it cf.} Sect.~\ref{I}). From a single-cell analysis similar to that in Sect.~\ref{sc2}, 
$\mu^{\Tr}$ is easily computed. One notes in particular that the distribution
of tracer energy is ${\rm const.} \sqrt x e^{-\beta x}$, while disk energy
is as before. A simple computation then gives $e=3s$.
\end{itemize}
These examples demonstrate clearly that the relation between $s$ and $e$
is entirely a function of the local structure. In the case of Hamiltonian
systems,
we see that it is also dimension-dependent.

\medskip
The interpretation of results in Sect.~\ref{interpret} applies to all of the models above.


\subsection{Ergodicity issues} 

Questions of ergodicity for the chains in Theorem~\ref{thm2} are beyond 
the scope of this paper. We include only brief discussions of the following
three aspects of the problem:

\medskip
\noindent {\it 1. Randomness in the injection process}

\smallskip
Among the various features of our models, the one the most
responsible for promoting ergodicity is
the randomness with which new tracers are injected into the system.
We observe, however, that this genuinely stochastic behavior occurs only 
at the two ends of the chain,  and even there, the transition probabilities 
do not have densities with respect to the underlying Lebesgue measure.
The problem is thus one of {\it controllability} involving the deterministic part
of the dynamics. 

\medskip
\noindent {\it 2. Hyperbolicity of billiard dynamics: a necessary 
condition}

\smallskip
Let $\Delta_N \subset {\mathbb R}^2$ denote the
playground for the tracers in the $N$-chain. That is to say, it is the
union of $N$ copies of $\Gamma$ arranged in the configuration 
shown in Fig.~\ref{f:cells} with open passages between adjacent copies
of $\Gamma$. The presence of one of more tracers being {\it trapped} 
in $\Delta_N$ without contact with any of the turning disks or the openings 
at the two ends (\ie, $\gammal^{(1)}$ and $\gammar^{(N)}$) is clearly 
an obstruction to ergodicity. This scenario is easily ruled out by choosing
$\Gamma_0$ to have concave (or scattering) walls. Such a choice
of $\Gamma_0$ implies that $\partial \Delta_N$ also has concave 
boundaries, and the free motion of a  particle in a domain with concave
boundaries is well known to be hyperbolic and ergodic  
\cite{Sinai1970,Liverani1995}.

We do not know if the absence of trapped tracers in the sense above
implies ergodicity.

\medskip
\noindent {\it 3. Enhancing ergodicity}

\smallskip Without (formally) guaranteeing ergodicity, 
various measures can be taken to ``enhance" it, meaning 
to make the system appear for practical purposes as close to being
ergodic as one wishes. For example, one can introduce more 
scattering within each cell by increasing the curvature of the
walls of $\Gamma_0$, or alternately, one could add convex bodies inside
$\Gamma_0$ that play the role of Lorentz scatterers. 
Another possibility is to add a small amount of noise, and 
a third is to increase the injection rates: physical  intuition says 
that the larger the number of tracers in the system, the more likely 
stored energy will behave ergodically.


\subsection{Results of simulations}\label{sim2}

To check the applicability of the theory proposed in
Sects.~\ref{fam}--\ref{prof} 
to real and finite systems, we have done extensive simulations some of
which we describe in this subsection. The domain $\Gamma_0$ used
in our simulations is as shown in Fig.~\ref{f:cells}. Actual specifications 
of $\Gamma_0$ are as follows: We start with a square of
sides $2$, subtracting from it first 4 disks of radius 1.15 centered
at the 4 corners of the square. Two openings corresponding to 
$\gammal$ and $\gammar$ are then created on the left and right;
each has length  0.02. This completes the definition of $\Gamma_0$. 
The disk $D$ is located at the center of the square; it has radius $r=0.0793$.

Our choice of domain was influenced by the following factors:
First, $\partial \Gamma_0$ is taken to be piecewise concave
to promote ergodicity. 
Second is the size of the disk: 
A disk that is too small is hit by a tracer 
only rarely; many tracers may pass through the cell without interacting 
with the disk (this is analogous to having a large $\delta$ in Sect.~\ref{rh}).
A disk that is too large (relative to the domain in which it can fit) 
may cause an unduly large fraction of tracers
entering the box to exit immediately from the same side. Both
scenarios lead to large time-correlations, which are well known to
impede the speed of convergence to $\mu_N$ in a finite chain.
They may also affect the infinite-volume limit.

We have found the geometry and specifications above to work
quite well, with a tracer making, on average, about 71 collisions 
while in a cell. Of these collisions, about 12.5 are with the disk.

For the single cell (with the geometry above) plugged to two identical
heat baths, we have tested the system extensively for ergodicity. To the degree
that one can ascertain from simulations, there is an ergodic component
covering nearly 100$\%$ of the phase space. The various energy 
distributions as well as the Poisson distribution of the number of tracers 
present agree perfectly with those predicted by Proposition~\ref{mu}.

Simulations for chains of 20 to 60 cells with the choice of $r$ and $|\gamma|$
above showed very good agreement with the theory. A sample of the fits for
$Q(\xi)$, $s(\xi)$ and $E(\xi)$ for $9\cdot 10^9$ events and 30 sites 
is shown in Figure 3. Here the bounds in Assumption 3 are very small, that is
to say, the ejection rates to the left and right are very close to $50/50$. 
We have also investigated the quantity $\alpha $ in Assumption 3 for various
values of $r$ and $|\gamma|$, up to $r=0.23$ (which is quite close to the 
maximum-size disk that can be fitted into the domain $\Gamma_0$) and 
$|\gamma|=0.06$. Our findings are consistent with the
discussion in Sect.~\ref{prof}.

In addition to these profiles, we have also verified
directly Assumption 2, which asserts that the distributions of energy and 
tracers within each cell are in accordance with those given by $\mu^{T,\j}$ 
for some $T, \j$ depending on the cell. A sample of these results is shown
in Fig.~\ref{fig:fig4}.
\begin{figure}[ht]
\C{\psfig{file=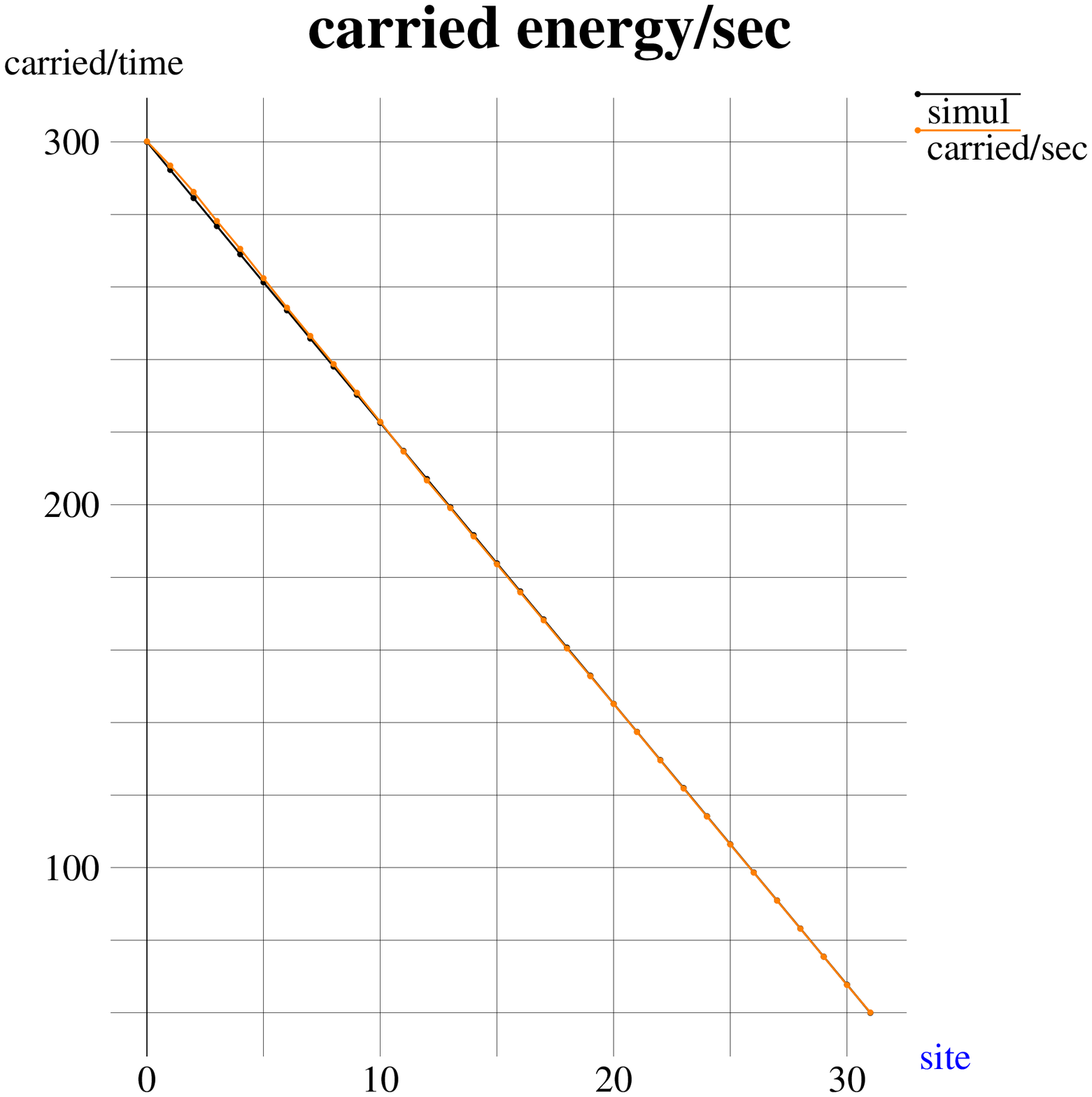,width=\x}\ \
\psfig{file=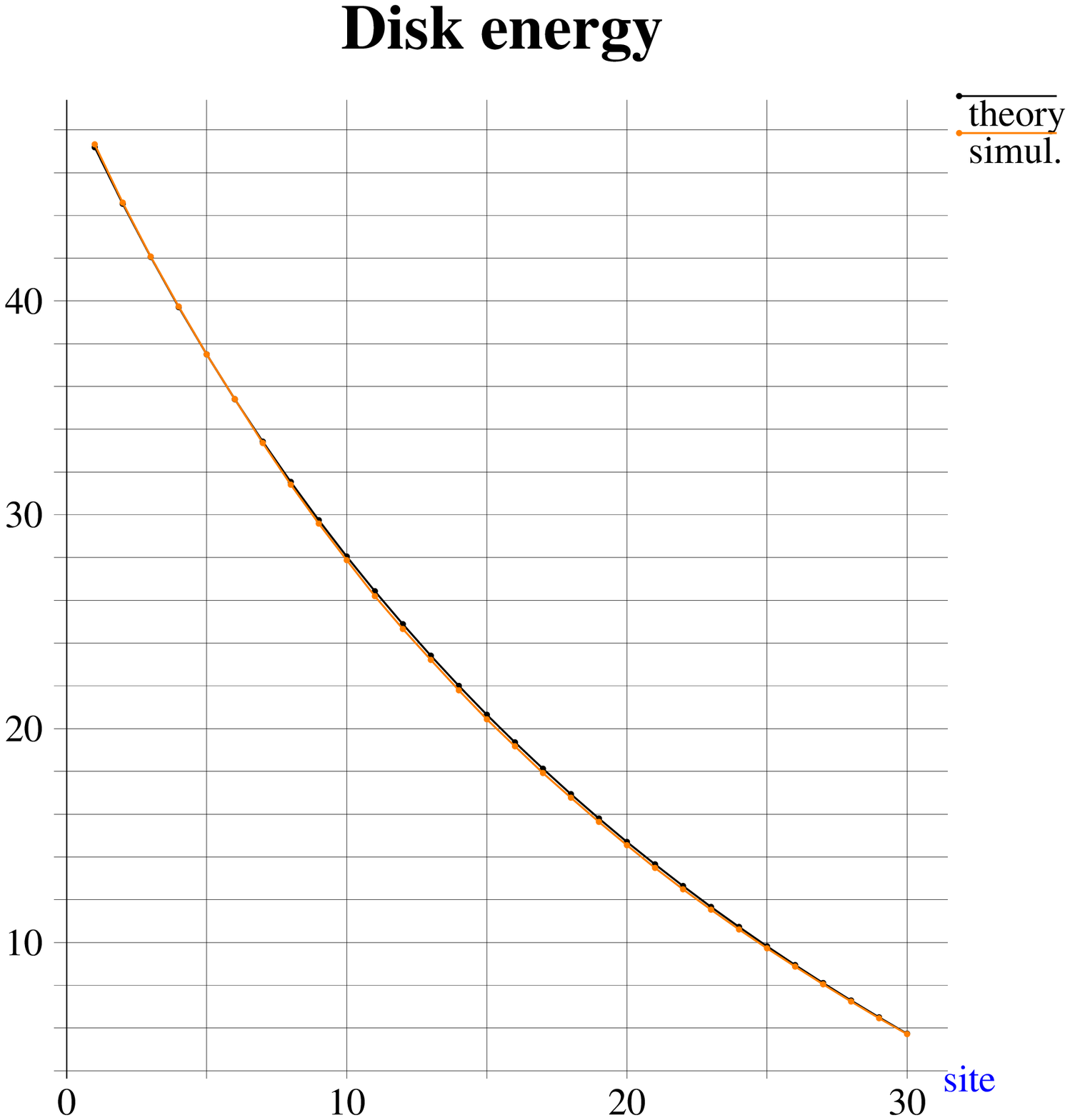,width=\x}} 
\C{\psfig{file=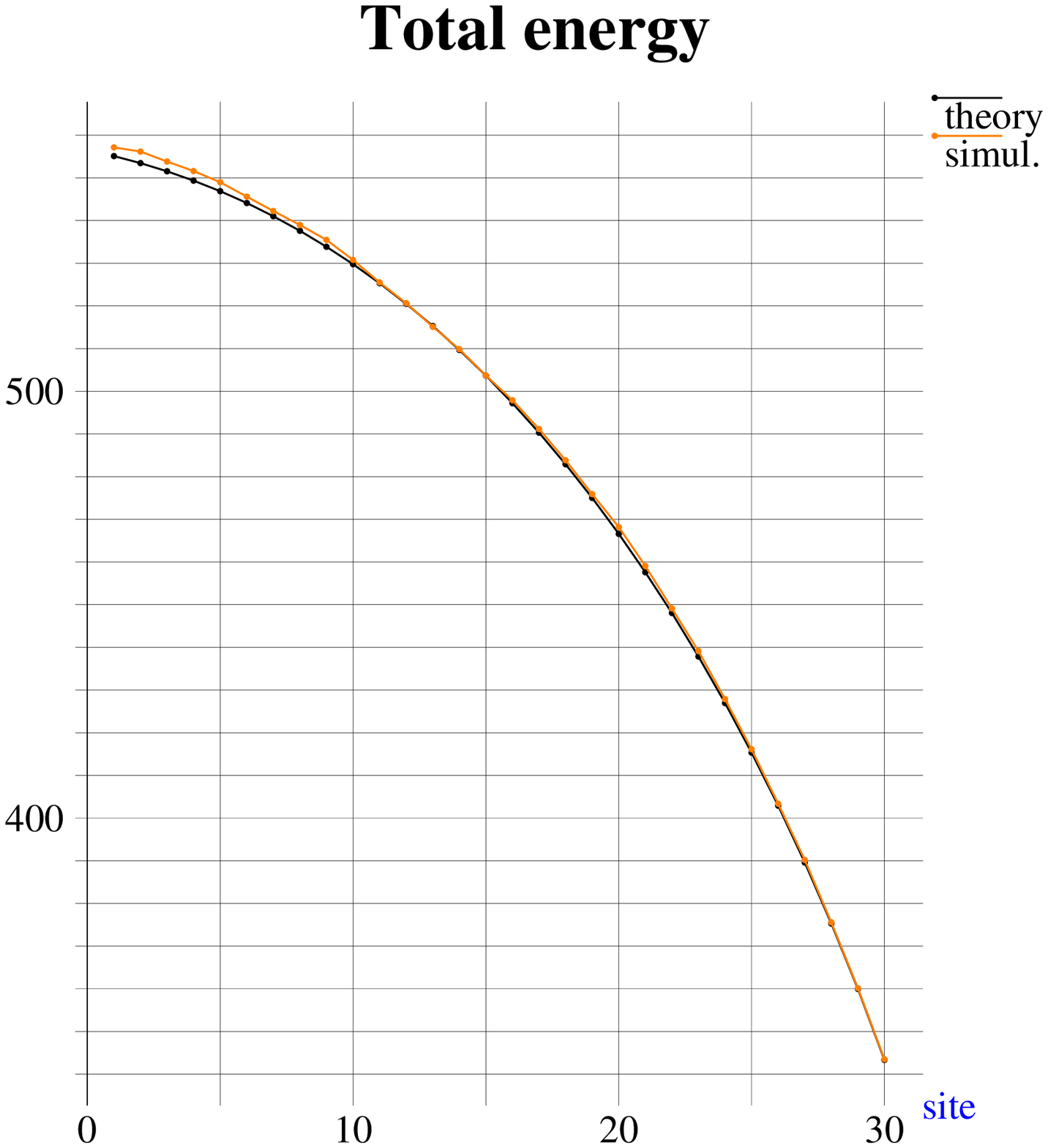,width=\x}}
\caption{{Rotating disks model with chain of 30 cells, temperatures
$\tl=100$, $\tr=10$, and injection rates  $\jl=1$, $\jr=2$.} 
Top left: $Q_i$, energy transported out of site $i$ per unit time
as a  function of $i$. Top right: Mean disk energy  $s_i$.
Bottom: Mean total energy $E_i$.}\label{fig:chain1}
\end{figure}
\begin{figure}
\C{
\psfig{file=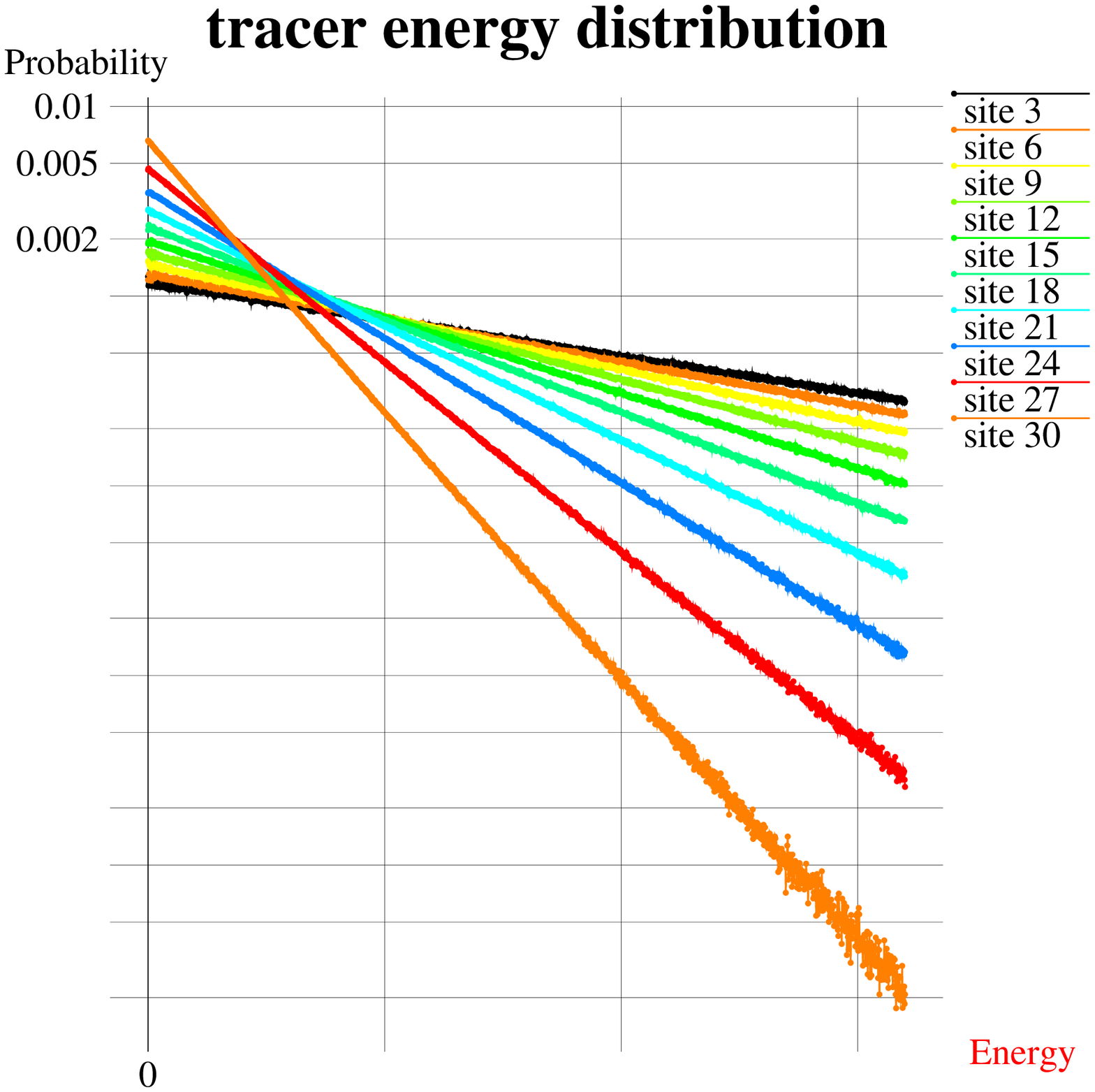,width=\x}
\psfig{file=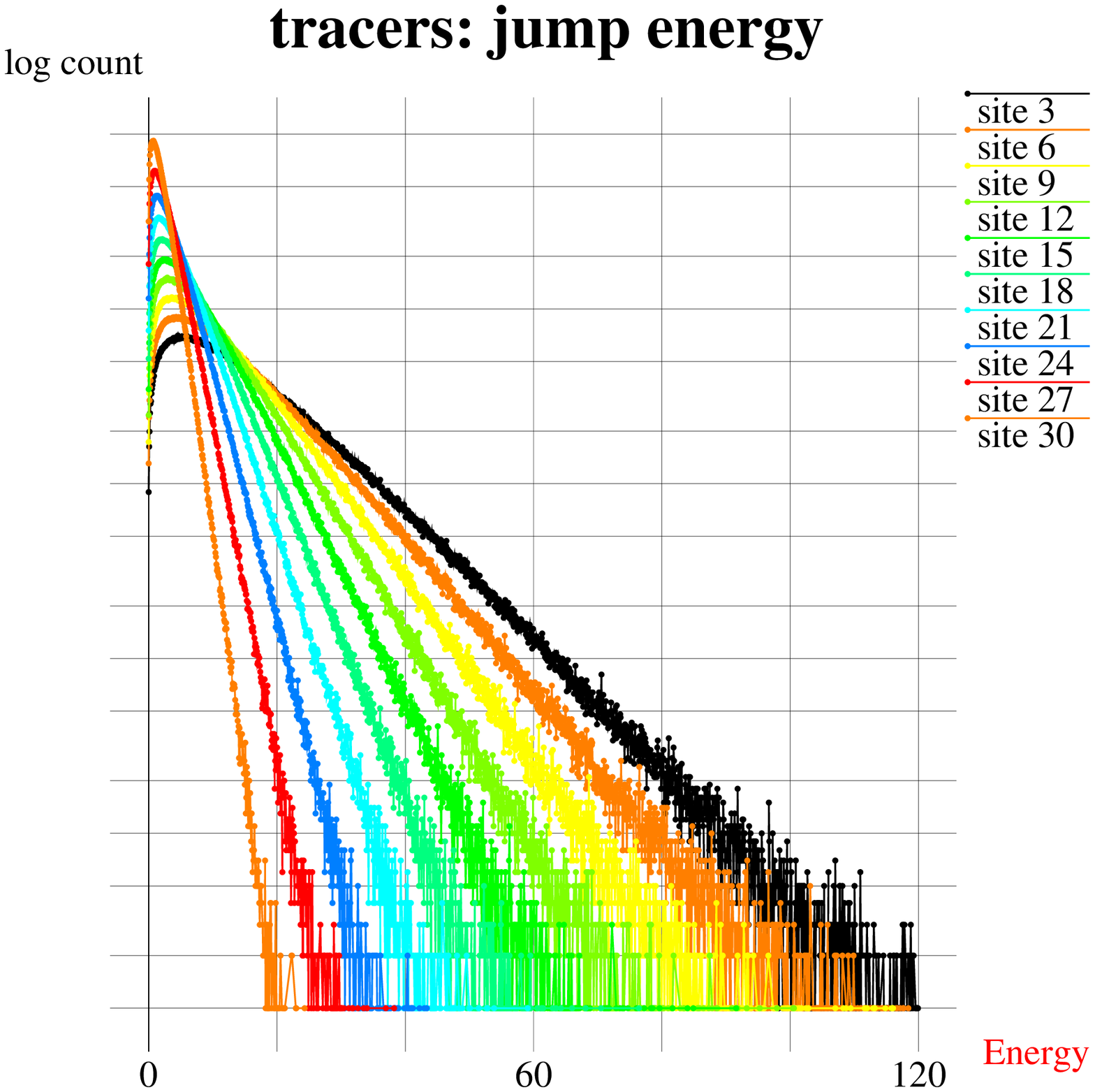,width=\x}} 
\C{\psfig{file=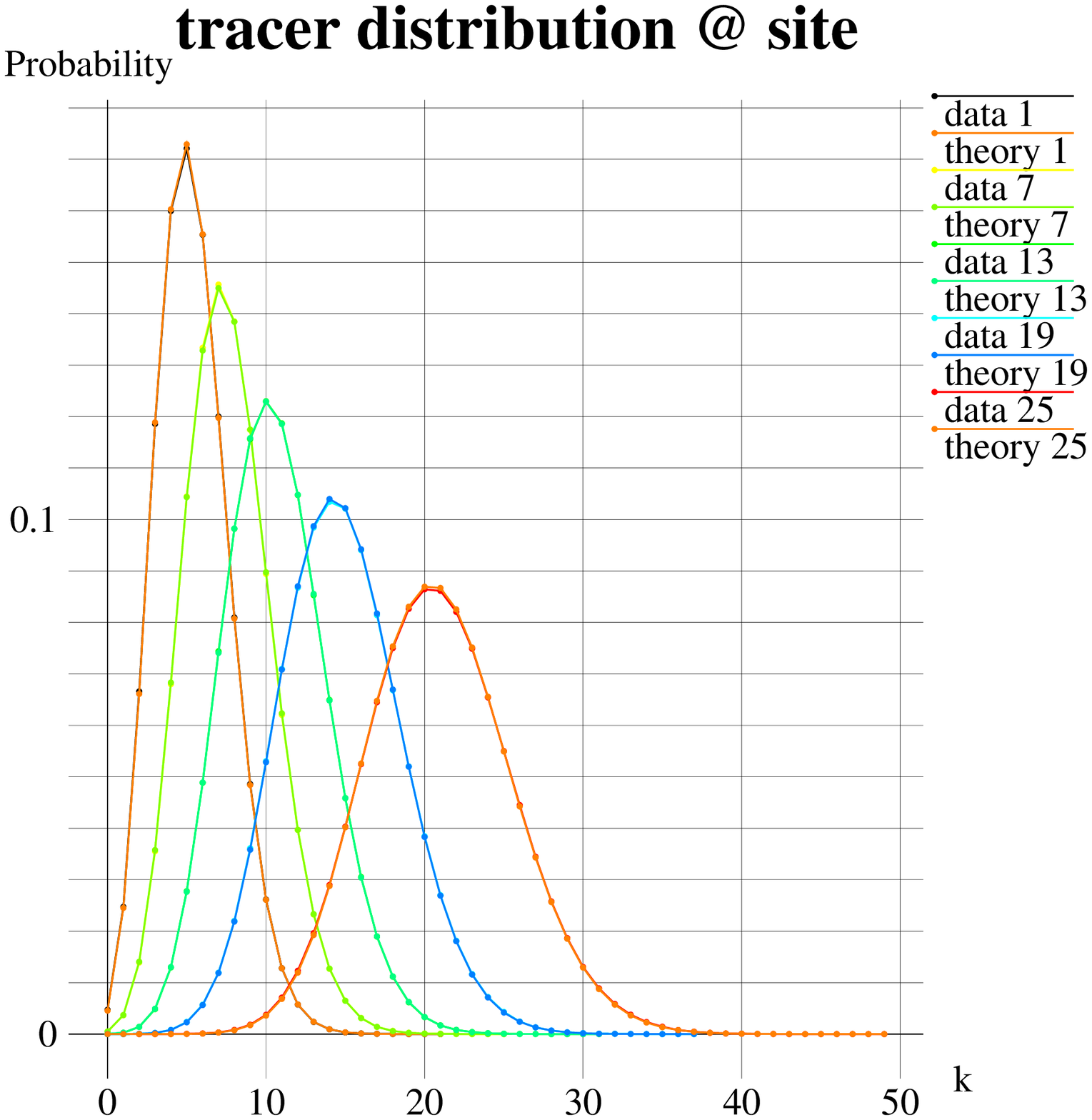,width=\x}}
\caption{Same parameters as in Fig.~\ref{fig:chain1}. 
Top 2 figures show semi-$\log$ plots of tracer energy distributions
at various sites. Top left: Densities of tracer energies inside boxes 
(theory predicts $ \beta e^{-\beta x}$).
Top right: Densities of tracer energies upon exiting the various boxes 
(theory predicts $2\beta^{\frac{3}{2}} \sqrt{x/ \pi} e^{-\beta x}$).
Bottom: Distribution of numbers of tracers at several sites
(theory predicts Poisson distribution).}\label{fig:fig4}
\end{figure}


\subsection{Related models}\label{concrete}

In this subsection we recall from the literature a few models that in their
original or slightly modified form can be regarded as approximate realizations
of the class described in Sect.~\ref{gsmodm} of this paper. For more
complete accounts, see the review papers
\cite{Reybellet2000,Lepri2003,Casati2003}.

The models which come closest to ours, and which to some degree 
inspired this work, are those in
\cite{Larralde2002,Larralde2003}. 
In these papers, the authors carried out a numerical study of
a system comprised of an array of disks similar to those in Sect.~\ref{fam} but
arranged in two rows with periodic boundaries (in the vertical direction). These disks interact
via tracers following the rules first used in \cite{Rateitschak2000}.
We have adopted the same local rules, but have elected to arrange our disks 
in a single row to simplify the analysis.

There is a number of papers dealing with mechanical gadgets that
on some level appear similar to ours. For example, in \cite{Li2004,Gruber2004},
vertical plates are pushed back and forth by particles trapped between them.
The main difference between these models and ours is that they have
 exactly one ``tracer'' in each ``cell". 
In this respect, these models are closer to our earlier work \cite{EY1}
in which locked-in tracers were considered. Ding-a-ling and ding-dong models
belong essentially to the same class \cite{Casati1984,Prosen1992,
Garrido2001,poschhoover1998}.

We mention that nonlinearities of profile are difficult to see 
when the temperature differences at the two ends are relatively small
(in fact, what counts in many cases, including the models studied in this
paper, is the {\it ratio} of temperatures at the two ends). This may 
explain why some authors have reported linear profiles when 
our analysis suggests that may not be the case.

Finally, we mention a very well-studied situation, namely that 
of the Fermi-Pasta-Ulam chain. In this model, and in many others, 
there is a potential of the form
$$
U(x_i -x_{i+1}) + V(x_i )~,
$$
with $U$ and $V$ functions that grow to $\infty$ and $x_i $ the
coordinates of a chain of anharmonic oscillators. The pinning potential $V$
plays the role of the ``tank" in our models, while the interparticle potential is
more akin to the role of the tracers. This class of models is
difficult to handle because in contrast to the basic setup in our study, 
there is no clear separation of the pinning and  interaction energies.

\bigskip 
\noindent{\bf Acknowledgments:} The authors thank O. Lanford
for helpful discussions. JPE acknowledges the Courant Institute, 
and LSY the University of Geneva, for their hospitality.

\bibliographystyle{JPE}
\markboth{\sc \refname}{\sc \refname}
\bibliography{refs}

\def\Rom#1{\uppercase\expandafter{\romannumeral #1}}\def\u#1{{\accent"15
  #1}}\def\cprime{$'$} \def\cprime{$'$}
\begin{thebibliography}{10}

\bibitem{Reybellet2000}
F.~Bonetto, J.~L. Lebowitz, and L.~Rey-Bellet.
\newblock Fourier's law: a challenge to theorists.
\newblock In: {\em Mathematical physics 2000\/} (London: Imp. Coll. Press,
  2000), pp. 128--150.

\bibitem{Casati1984}
G.~Casati, J.~Ford, F.~Vivaldi, and W.~Visscher.
\newblock One-dimensional classical many-body system having a normal thermal
  conduction.
\newblock {\em Phys. Rev. Lett.\/} {\bf 52} (1984), 1861--1864.

\bibitem{Groot1962}
S.~De~Groot and P.~Mazur.
\newblock {\em Non-Equilibrium Thermodynamics\/} (North Holland, 1962).

\bibitem{Demasi1991}
A.~De~Masi and E.~Presutti.
\newblock {\em Mathematical methods for hydrodynamic limits\/}, volume 1501 of
  {\em Lecture Notes in Mathematics\/} (Berlin: Springer-Verlag, 1991).

\bibitem{EY1}
J.-P. Eckmann and L.-S. Young.
\newblock Temperature profiles in {H}amiltonian heat conduction.
\newblock {\em Europhysics Letters\/} {\bf 68} (2004), 790--796.

\bibitem{Garrido2001}
P.~Garrido, P.~Hurtado, and B.~Nadrowski.
\newblock Simple one-dimensional model of heat conduction which obeys fourier's
  law.
\newblock {\em Phys. Rev. Lett.\/} {\bf 86} (2001), 5486--5489.

\bibitem{Gruber2004}
C.~Gruber and A.~Lesne.
\newblock Hamiltonian model of heat conductivity and {F}ourier law.
\newblock {\em Preprint\/} .

\bibitem{Kipnis1999}
C.~Kipnis and C.~Landim.
\newblock {\em Scaling limits of interacting particle systems\/}, volume 320 of
  {\em Grundlehren der Mathematischen Wissenschaften [Fundamental Principles of
  Mathematical Sciences]\/} (Berlin: Springer-Verlag, 1999).

\bibitem{Larralde2003}
H.~Larralde, F.~Leyvraz, and C.~Mej\'\i{a}-Monasterio.
\newblock Transport properties of a modified {L}orentz gas.
\newblock {\em J. Stat. Phys.\/} {\bf 113} (2003), 197--231.

\bibitem{Lepri2003}
S.~Lepri, R.~Livi, and A.~Politi.
\newblock Thermal conduction in classical low-dimensional lattices.
\newblock {\em Phys. Rep.\/} {\bf 377} (2003), 1--80.

\bibitem{Li2004}
B.~Li, G.~Casati, J.~Wang, and T.~Prosen.
\newblock Fourier law in the alternate mass hard-core potential chain.
\newblock {\em Phys. Rev. Lett.\/} {\bf 92} (2004), 254301.

\bibitem{Casati2003}
B.~Li, G.~Casati, J.~Wang, and T.~Prosen.
\newblock Fourier law in the alternate mass hard-core potential chain
  (cond-mat/0307692).

\bibitem{Liverani1995}
C.~Liverani and M.~P. Wojtkowski.
\newblock Ergodicity in {H}amiltonian systems.
\newblock In: {\em Dynamics reported\/}, volume~4 of {\em Dynam. Report.
  Expositions Dynam. Systems (N.S.)\/} (Berlin: Springer, 1995), pp. 130--202.

\bibitem{Larralde2002}
C.~Mej\'\i{a}-Monasterio, H.~Larralde, and F.~Leyvraz.
\newblock Coupled normal heat and matter transport in a simple model system.
\newblock {\em Phys. Rev. Lett.\/} {\bf 86} (2001), 5417--5420.

\bibitem{poschhoover1998}
H.~A. Posch and W.~G. Hoover.
\newblock Heat conduction in one-dimensional chains and nonequilibrium
  {L}yapunov spctrum.
\newblock {\em Phys. Rev. E\/} {\bf 58} (1998), 4344--4350.

\bibitem{Prosen1992}
T.~Prosen and M.~Robnik.
\newblock Energy transport and detailed verification of fourier heat law in a
  chain of colliding harmonic oscillators.
\newblock {\em J. Physics. A\/} {\bf 25} (1992), 3449--3478.

\bibitem{Rateitschak2000}
K.~Rateitschak, R.~Klages, and G.~Nicolis.
\newblock Thermostating by deterministic scattering: the periodic {L}orentz
  gas,.
\newblock {\em J. Stat. Phys.\/} {\bf 99} (2000), 1339--1364.

\bibitem{Sinai1970}
J.~G. Sina{\u\i}.
\newblock Dynamical systems with elastic reflections. {E}rgodic properties of
  dispersing billiards.
\newblock {\em Uspehi Mat. Nauk\/} {\bf 25} (1970), 141--192.

\bibitem{Spohn1991}
H.~Spohn.
\newblock {\em Large Scale Dynamics of Interacting Particles\/}.
\newblock Texts and Monographs in Physics (Heidelberg: Springer-Verlag, 1991).

\end{thebibliography}

\end{document}